\documentclass[a4paper,11pt]{article}
\pdfoutput=1 % if your are submitting a pdflate\xi (i.e. if you have
\usepackage[T1]{fontenc} % if needed
\usepackage{enumitem}

\usepackage{tikz}
\usetikzlibrary{decorations.pathmorphing}
\usepackage{subcaption} 
\usepackage{caption}
\usepackage{tkz-tab}
\usepackage{amsmath}
\usepackage{amssymb}
\usepackage{amsfonts}
\usepackage{amsthm}
\usepackage{mathrsfs}
\usepackage{comment}
\usepackage{graphicx}
\usepackage{epstopdf}
\usepackage{empheq}
\usepackage{hyperref}

\def\be {\begin{equation}}
\def\ee {\end{equation}}
\def\bea {\begin{eqnarray}}
\def\eea {\end{eqnarray}}
\def\nn {\nonumber}
\def\cL{{\cal L}}

\newcommand{\mysection}[1]{\section{#1}\setcounter{equation}{0}}

\definecolor{purple}{rgb}{0.7,0,1}

\begin{document}

\begin{titlepage}

%\begin{flushright}
%ArXiv: NNNN.NNNNN [gr-qc]\\
%UB-NNNN
%\end{flushright}
%\vskip 1cm

\vspace{2cm}

\begin{center}
{\bf\Large Self-similarity in Einstein-Maxwell-dilaton theories\\
and critical collapse}
\end{center}

%\begin{center}\rule[0.1in]{13cm}{0.3mm} \end{center}

\vspace{0.5cm}

\begin{center}
{\large Jorge V. Rocha$^1$ and Marija Toma\v{s}evi\'c$^1$}
\bigskip\bigskip

\vspace{0.8cm} { \normalsize
$^1$Departament de F\'isica Qu\`antica i Astrof\'isica and Institut de Ci\`encies del Cosmos (ICCUB),\\
Universitat de Barcelona, Mart\'i i Franqu\`es 1, E-08028 Barcelona, Spain\\

\vspace*{0.2cm}

\texttt{jvrocha@icc.ub.edu}\; %\note{Corresponding author.}
and\;
\texttt{mtomasevic@icc.ub.edu} }

\end{center}

\vspace{1cm}

\abstract{
We study continuously self-similar solutions of four-dimensional Einstein-Maxwell-dilaton theory, with an arbitrary dilaton coupling. %and the inclusion of a potential for the scalar field.
Self-similarity is an emergent symmetry of gravitational collapse near the threshold of black hole formation. The resulting `critical collapse' picture has been intensively studied in the past for self-gravitating scalar fields or perfect fluids, but little is known concerning other systems.
Here we assess the impact of gauge fields on critical collapse, in the context of low-energy string theories.

Matter fields need not inherit the symmetries of a spacetime. We determine the homothetic conditions that scale-invariance of the metric imposes on the dilaton and electromagnetic fields, and we obtain their general solution. The inclusion of a potential for the dilaton is compatible with the homothetic conditions if and only if it is of the Liouville type.

By imposing also spherical symmetry, a detailed analysis of critical collapse in these systems is possible by casting the field equations as an autonomous system. We find analytically that Choptuik's critical exponent depends on the dilaton coupling. Despite this and the presence of two novel fixed points, the electromagnetic field necessarily vanishes for the critical solution.
}

\end{titlepage}

%\maketitle
%\flushbottom

%%%%%%%%%%%%%%%%%%%%%%%%%%%%%%%%%%%%%%%%%%
\mysection{Introduction}\label{sec:Intro}
%%%%%%%%%%%%%%%%%%%%%%%%%%%%%%%%%%%%%%%%%%

Critical behavior plays a prominent role in gravitational collapse~\cite{Gundlach:1999cu}.
As with all critical phenomena, its importance stems from its universality properties, which entails one can predict general features of the outcome of a process of gravitational collapse irrespective of details of initial data. This is most clearly illustrated by the universal nature of the critical exponent controlling how the mass of the black hole formed, $M$, approaches the threshold value $M_0$ (typically zero) as any given parameter $p$ determining the initial conditions is tuned to criticality ($p=p^*$),
\be
M-M_0 \sim (p-p^*)^\beta\,.
\label{eq:Choptuik}
\ee
This famous Choptuik scaling law~\cite{Choptuik:1992jv} is a hallmark of critical collapse, and the critical exponent $\beta$ is universal in the sense that it is the same for all families of initial data parametrized by a single parameter $p$.

Critical collapse has been extensively studied in the context of a minimally coupled scalar field~\cite{Choptuik:1992jv,Hamade:1995ce,Gundlach:1996eg}, which was the original arena for Choptuik's seminal studies, as well as for self-gravitating perfect fluids~\cite{Evans:1994pj,Koike:1995jm,Maison:1995cc,Neilsen:1998qc,Koike:1999eg,Brady:2002iz}. Both classes of investigations build on mathematical results by Christodoulou~\cite{Christodoulou:1986zr,Christodoulou:1984mz} and are thus typically performed under the simplifying assumption of spherical symmetry\footnote{However, see~\cite{Abrahams:1993wa,Baumgarte:2016xjw,Gundlach:2017tqq} for similar investigations away from spherical symmetry.}.
Another key feature in critical collapse is the emergence of self-similarity as an intermediate attractor for near-critical solutions~\cite{Gundlach:1999cu,Carr:1998at,Musco:2012au}. This feature comes in two possible flavors, continuous or discrete, and it is not understood how to tell {\it a priori} which one will be revealed by any given system. For instance, the minimally coupled massless scalar field exhibits {\em discrete} self-similarity, whereas gravitational collapse of perfect fluids reveals {\em continuous} self-similarity.

Much less attention has been dedicated to critical collapse in alternative theories of gravity, such as those arising in low-energy string theory. In this paper we investigate continuous self-similar collapse in Einstein-Maxwell-dilaton (EMD) theories. This is a well-motivated model that captures the main features of four-dimensional effective string theories\footnote{Actually, for certain choices of the dilaton coupling constant it is known to be a consistent truncation of string theory, as mentioned in the following.}, but it can also be regarded as an extension of the standard Einstein-massless scalar system.
It might seem that solutions at the threshold of black hole formation could be largely discounted as 'non-generic' configurations, but they are very important in at least one respect: they effectively correspond to the formation of naked singularities and therefore have some bearing on the cosmic censorship conjecture~\cite{Goldwirth:1987nu}. In particular, tuning initial data near criticality provides the means to probe the deep quantum gravity regime. In the present context this would be the string theory providing the ultraviolet completion for the EMD model.

The dynamics of black hole mergers within this theory was examined only recently~\cite{Hirschmann:2017psw}, indicating such processes are essentially indistinguishable from those in general relativity (GR) ---and therefore compatible with LIGO-VIRGO detections~\cite{TheLIGOScientific:2016src}, thus promoting EMD to an  interesting alternative to GR--- when the charges are small. Nevertheless, the presence of a scalar field coupled not only to gravity, but to a gauge field as well, advises the use of caution when considering such theories as viable alternatives to GR. For instance, the relaxation of a linearly perturbed EMD black hole in isolation has been shown to lead to significant departures from Einstein-Maxwell theory in the electromagnetic sector~\cite{Brito:2018hjh}.

Somewhat surprisingly, no study of gravitational collapse has been conducted in EMD theory so far.  Perhaps the closest related investigation is the one by Hamad\'e {\it et al.}~\cite{Hamade:1995jx}, which studies the gravitationally coupled axion-dilaton system (see also~\cite{AlvarezGaume:2011rk}). Both this system and the EMD model are consistent truncations of the 4D low-energy heterotic string, but they are distinct: the former does not include the Maxwell field, whereas here we consider a single real scalar, the dilaton. In any case, Ref.~\cite{Hamade:1995jx} showed numerically that the critical solution at the threshold of black hole formation is continuously self-similar, thus suggesting that the same assumption for critical collapse in the EMD system is justified.

In the interest of generality, we allow the dilaton coupling parameter $a$ to take arbitrary values. Furthermore, we will extend the pure EMD system by including a Liouville potential for the scalar field.
Potentials of this type contain as a particular case a simple cosmological constant, and they arise naturally in string effective actions, either through symmetry breaking mechanisms~\cite{Poletti:1994ff} or by dimensional reduction of a parent cosmological constant term~\cite{Charmousis:2009xr}.
Interestingly, this class of dilaton potentials is consistent with continuous self-similarity, as we will see.

In this paper we shall consider exclusively source-free solutions of the EMD field equations.
Moreover, the \textit{entire} family of solutions we present in Section~\ref{sec:SphericalCSS} ---and not just the critical solution--- displays continuous self-similarity. Hence, in this respect our analysis differs from the usual considerations of gravitational critical collapse employing numerical simulations, where self-similarity arises only at the threshold between full dispersal (or formation of a \textit{star}-like solution) and collapse to a black hole. Instead, we partly follow previous work by Brady on homothetic scalar field collapse~\cite{Brady:1994aq}, which heavily relies on the use of continuous self-similarity (see~\cite{Christodoulou:1994hg} for earlier work along similar lines). Accordingly, our critical solutions interpolate between black holes and \textit{naked singularities}.

The assumption of spherical symmetry and continuous self-similarity appear quite restrictive, so let us briefly discuss their validity. Spherical symmetry has always played an important role in gravitation. Here we note that it is a good approximation when considering collapsing matter close to criticality, since it becomes more accurate at the later stages when the system relaxes down to a stable state~\cite{Baumgarte:2016xjw} (see however Ref.~\cite{Gundlach:2017tqq} for an indication that this picture might change at very large rotation rates).
On the other hand, self-similarity arises naturally as one approaches the threshold of black hole formation, as we already alluded to. For the case of a massless scalar field, Choptuik showed~\cite{Choptuik:1992jv} that near the critical solution, (discrete) self-similarity emerges in the form of ``echoes''. It can also be mentioned that the similarity hypothesis, concerning the tendency of gravitational systems to evolve to self-similar form and thus to asymptotically approach a more symmetric state, has been analyzed in depth~\cite{Carr:1998at}, including in particular the case of spherically symmetric solutions.

%%%%%%
\paragraph{The continuous self-similarity property.}
%%%%%%
The assumption of continuous self-similarity (CSS) entails a significant degree of simplification. In fact, it is as powerful as an isometry; the only difference is that for the former there exists a diffeomorphism that leaves the metric invariant only up to a constant factor. In technical terms, the spacetime possesses a homothetic vector field (HVF) $\xi$ such that~\cite{Cahill:1970ew,Carr:1998at}
\be
\cL_\xi g_{\mu\nu} = 2 g_{\mu\nu}\,,
\label{eq:SScond}
\ee
where $g_{\mu\nu}$ stands for the metric tensor.
In practice, CSS imposes a drastic simplification when in combination with spherical symmetry: solutions depend on a single coordinate and the equations of motion reduce to ODEs.
The present article focuses on this kind of self-similarity, so we will have nothing to say about discrete self-similarity.

In a CSS spacetime, as defined above, there is total absence of a characteristic scale. Hence, at first sight this property appears to be equivalent to scale-invariance. While this is true for the gravitational sector (metric functions depend only on ratios of coordinates~\cite{Cahill:1970ew}), this symmetry is {\em not} necessarily inherited by matter fields coupled to gravity~\cite{Carr:1998at,Smolic:2015txa}.
In fact, a minimally coupled scalar field $\Phi$ need not be invariant along the homothetic vector field, and more generally one has~\cite{Brady:1994aq}, as re-derived in Section~\ref{sec:HomoConds} below,
\be
\cL_\xi \Phi = -\kappa\,,
\label{eq:homo_dilaton}
\ee
where $\kappa$ is an arbitrary real number. (The minus sign is conventional.)
In the case of complex scalar fields, the occurring global internal symmetries can mix up with spacetime symmetries and other homothetic actions are possible~\cite{Hirschmann:1994du,Eardley:1995ns}.

On the other hand, for the Einstein-Maxwell system the self-similarity condition~\eqref{eq:SScond} implies~\cite{Wainwright:1976uu, Wainwright:1976ux}
\be
\cL_\xi F_{\mu\nu} = F_{\mu\nu} + \widetilde{\kappa} \star\! F_{\mu\nu}\,,
\label{eq:homo_Maxwell}
\ee
where $F_{\mu\nu}$ denotes the Maxwell field strength, $\star F_{\mu\nu}$ is its Hodge dual, and $\widetilde{\kappa}$ is an undetermined scalar quantity\footnote{The scalar $\widetilde{\kappa}$ must be constant in the case of a non-null electromagnetic field. This condition is somewhat relaxed when the electromagnetic field is null, characterized by $F_{\mu\nu}F^{\mu\nu}=0=F_{\mu\nu}\star F^{\mu\nu}$. In that case, $k_{[\mu}\nabla_{\nu]}\widetilde{\kappa}=0$, where $k^\mu$ is the repeated principal null direction of $F_{\mu\nu}$ ~\cite{Wainwright:1976ux}.} that indicates a departure of the electromagnetic field strength from inheriting the homothety of the metric.

%%%%%%
\paragraph{Overview of main results.}
%%%%%%
The primary questions we address in the present study are twofold: (i) What are the conditions that continuous self-similarity dictates for the dilaton and Maxwell fields; (ii) Does the inclusion of the electromagnetic field (and possibly a potential for the dilaton) alter the critical collapse picture obtained for the spherical Einstein-massless scalar system? In particular, does it affect the critical exponent $\beta$, which governs Choptuik scaling close to the threshold of black hole formation?
%(iii) More generally, do there exist `regular' (spherically symmetric) CSS solutions in EMD with a non-vanishing electromagnetic field?

Concerning question (i), we will show that, for the combined Einstein-Maxwell-dilaton system, the self-similarity condition~\eqref{eq:SScond} allows slightly more general actions of the homothety on the dilaton and electromagnetic field strength.
The transformation law for the scalar field is still the same as~\eqref{eq:homo_dilaton}, but the coupling between the dilaton and the Maxwell field introduces modifications to Eq.~\eqref{eq:homo_Maxwell}.
The inclusion of a potential of the exponential (Liouville) type is consistent with continuous self-similarity, but only if the coefficient in the exponent of the potential is fixed in terms of the dilaton homothetic parameter $\kappa$ [see Eq.\eqref{eq:kappabeta}].

As for question (ii), we first note that consistency between the `time' and `radial'  components of the Maxwell equations demands that either the electric field vanishes or that the dilaton homothetic parameter $\kappa$ and the dilaton coupling $a$ are not independent [see Eq.\eqref{eq:akappa}].
Taking this constraint into account, the outcome of our analysis is the following:  in spherical symmetry, continuous self-similar vacuum solutions of the EMD system with a regular origin necessarily have vanishing electric field.  {\em In this sense, the electric field is irrelevant for critical collapse in the context of source-free EMD theories.} However, the critical exponent strongly depends on the homothetic parameter $\kappa$. For CSS solutions with a non-trivial electric field (obtained by relaxing the condition of regularity at the origin) the critical exponent is not the same for all EMD theories, given that $\kappa$ is fixed by the value of the dilaton coupling.  {\em Thus, the critical solution is not universal in the broader sense, as it depends on the dilaton coupling that selects a given theory within the whole EMD family. %of not being common to the whole family of EMD theories.
}

%%%%%%
\paragraph{Related literature.}
%%%%%%
There have been a few studies, both semi-analytic~\cite{Gundlach:1996vv} and fully numerical~\cite{Hod:1996ar,Oren:2003gp}, of gravitational collapse with charged scalar fields. The outcome was that the addition of electromagnetic charge does not change the (mass) critical exponent $\beta$. However, the dynamics displayed in this case differs from the Einstein-Maxwell-dilaton system since the governing equations of motion are distinct: while in the former theory the scalar field is derivatively coupled to the gauge vector potential, in the latter it couples directly to the field strength (squared) with a characteristic exponential form.

In this respect, reference~\cite{Borkowska:2011zh} is of more relevance for our purposes, since it investigated gravitational collapse in low-energy string theory, employing numerical evolutions. Unfortunately, this work addresses critical collapse only tangentially and offers no insight as to the appearance (or not) of self-similarity near criticality. Moreover, the system considered therein was coupled to an additional complex charged scalar field, which is absent in our study.

Some analytic solutions describing self-similar collapse in dilaton gravity were presented over the years, e.g.  \cite{Roberts:1989sk,Brady:1994xfa,deOliveira:1995cn,Yazadjiev:2003bp}, although none has been obtained for EMD theory. Ref.~\cite{Zhang:2014dfa} also derived exact collapsing solutions for the Einstein-scalar field system sharing some features akin to critical behavior, though not displaying any form of self-similarity.

%%%%%%
\paragraph{Outline of the paper.}
%%%%%%
The rest of the paper is organized as follows.
In the next section we present the family of EMD theories under consideration and its governing field equations.
The homothetic conditions imposed on the matter fields to be consistent with the self-similarity of the metric are derived and solved in Section~\ref{sec:SSconditions}. These results are used as inputs for the CSS collapses studied in the rest of the paper, but they can also be regarded as ans\"atze. So readers interested in quickly getting to the critical collapse analysis can skip Section~\ref{sec:SSconditions}  entirely.
In Section~\ref{sec:SphericalCSS} we restrict to continuous self-similar collapses in spherically symmetry. The equations of motion are cast in the form of an autonomous system, for which solutions are obtained as integral curves. The critical exponent is then extracted from the growing mode determined by linear perturbations around a fixed point of the dynamical system.
Section~\ref{sec:Conclusion} is devoted to discussion and outlook.
The article is also supplemented with some appendices. Appendices~\ref{sec:AppA} and~\ref{sec:AppB} collect several identities and a proof concerning homothetic vector fields and its action on the stress-energy tensor. Appendices~\ref{sec:AppC} and~\ref{sec:AppD} extend the study of Section~\ref{sec:SphericalCSS} to include a Liouville potential and for the case of a purely magnetic Maxwell field, respectively.

\bigskip

%%%%%%%%%%%%%%%%%%%%%%%%%%%%%%%%%%%%%%%%%%
\mysection{The Einstein-Maxwell-dilaton system}\label{sec:EOMs}
%%%%%%%%%%%%%%%%%%%%%%%%%%%%%%%%%%%%%%%%%%

The four-dimensional Einstein-Maxwell-dilaton model we consider is governed by the following action ($G=c=1$),
\be
{\cal S} = \frac{1}{16\pi} \int dx^4\, \sqrt{|g|} \left[R-2(\nabla\Phi)^2-e^{-2 a\Phi}F^2 -4V(\Phi) \right] \,.
\label{eq:action}
\ee
Here $g$ represents the determinant of the metric $g_{\mu\nu}$, $A_\mu$ is the Maxwell field whose field strength is $F_{\mu\nu}=\partial_\mu A_\nu-\partial_\nu A_\mu$, and $\Phi$ is a scalar field, the dilaton, for which we include a generic potential $V(\Phi)$, at this point. Later on we will restrict $V$ to be of the Liouville type. The Ricci tensor, denoted by $R_{\mu\nu}$, yields the curvature scalar $R$ upon contraction. The scalar and vector fields couple with a strength controlled by the so-called dilaton coupling constant $a$. Special values of the dilaton coupling appear naturally in different contexts. The four-dimensional low-energy effective action for heterotic string theory takes the form~\eqref{eq:action} with $a=1$, while $a=\sqrt{3}$ corresponds to Kaluza-Klein reduction of 5D Einstein gravity on the circle. Einstein-Maxwell theory is recovered by choosing $a=0, V(\Phi)=0$, and consistently setting the dilaton to zero.

The field equations derived from Eqs.~\eqref{eq:action} read 
\begin{subequations}\label{eq:fieldeqs}
\bea
&& \nabla^2 \Phi + \frac{a}{2}e^{-2 a\Phi} F_{\mu \nu} F^{\mu \nu} - \frac{dV}{d\Phi} = 0\,,\label{eq:dilatonEOM}\\
&& \nabla_\mu \left( e^{-2 a \Phi} F^{\mu \nu} \right) = 0\,,\label{eq:MaxwellEOM}\\
&& R_{\mu\nu} - \frac{1}{2}R g_{\mu\nu} = 8\pi  T_{\mu\nu} \equiv 8\pi \left(T_{\mu\nu}^{\rm (dil)} + T_{\mu\nu}^{\rm (EM)} \right)\,.\label{eq:EinsteinEOM}
\eea
\label{eq:EOM}
\end{subequations}
\!\!\!The full stress-energy tensor $T_{\mu\nu}$ has contributions from the dilaton and the electromagnetic field,
\begin{subequations}
\bea
8\pi T_{\mu\nu}^{\rm (dil)} &=& 2\nabla_\mu\Phi\nabla_\nu\Phi - g_{\mu\nu}\left[(\nabla\Phi)^2 + 2V(\Phi)\right]\,, \label{eq:Tdil}\\
8\pi T_{\mu\nu}^{\rm (EM)} &=& e^{-2a\Phi} \left( 2F_{\mu \sigma} {F_\nu}^\sigma -\frac{1}{2}g_{\mu \nu}F^2 \right)\,.
\eea
\end{subequations}
Generically, both the dilaton and Maxwell fields source the Einstein equations.
We refer to solutions without additional sources as `source-free' or simply `vacuum' solutions.

In four dimensions the electromagnetic stress-energy tensor is traceless (i.e., EM is conformal in 4D), $g^{\mu\nu} T_{\mu\nu}^{\rm (EM)}=0$, and therefore by contracting the Einstein equation~\eqref{eq:EinsteinEOM} with the inverse metric one obtains
\be
R = 2 \left(\nabla \Phi\right)^2 + 8V(\Phi)\,.
\label{eq:EinsteinScalar}
\ee
Replacing this back in the Einstein equation leads to
\be
R_{\mu\nu} = 2\nabla_\mu\Phi\nabla_\nu\Phi + 2g_{\mu\nu}V(\Phi) + e^{-2a\Phi} \left( 2F_{\mu \sigma} {F_\nu}^\sigma -\frac{1}{2}g_{\mu \nu}F^2 \right)\,.
\ee
%

%As a side remark, we note that the following identity is obtained by taking the divergence of Maxwell's equation~\eqref{eq:MaxwellEOM}:
%%
%\be
%(\nabla_\mu \Phi) \nabla_\nu F^{\mu\nu} = 0\,.
%\ee
%%

Exact source-free solutions of EMD in the static, spherically symmetric case (with vanishing potential) have been known for some time~\cite{Gibbons:1987ps,Garfinkle:1990qj,Kallosh:1992ii,Rasheed:1995zv}\footnote{Other solutions with unusual asymptotics have been presented in~\cite{Chan:1995fr,Charmousis:2009xr}.
%Ref.~\cite{Poletti:1994ww} numerically obtained static, charged dilatonic black holes with a negative cosmological constant.
%More general dilaton black holes are discussed in~\cite{Rakhmanov:1993yd}, but they are singular at the outer horizon. 
}.
Beyond this, but still restricting to spherically symmetric configurations, some time-dependent solutions are known analytically but, to the best of our knowledge, only when the dilaton coupling takes the heterotic string value~\cite{Gueven:1996zm,Aniceto:2017gtx,Lu:2014eta}. None of those solutions display continuous self-similarity.

%%%%%%%%%%%%%%%%%%%%%%%%%%%%%%%%%%%%%%%%%%
\mysection{Self-similarity conditions for the source-free EMD system\label{sec:SSconditions}}
%%%%%%%%%%%%%%%%%%%%%%%%%%%%%%%%%%%%%%%%%%

We restrict our investigations to self-similarity of the continuous kind.
As mentioned in the introduction, this amounts to assuming the existence of a homothetic vector field $\xi$ satisfying
\be
\cL_\xi g_{\mu\nu} = \nabla_\mu \xi_\nu + \nabla_\nu \xi_\mu = 2 g_{\mu\nu}\,.
\label{eq:SScondRepeat}
\ee
From this expression it follows a slew of relations involving the action of the homothety on tensors derived from the metric, as well as on matter fields, which are collected in Appendix~\ref{sec:AppA}.

Our goal now is to determine ---by applying the Lie derivative along a HVF to the field equations--- what is the action of $\cL_\xi$ on the dilaton and Maxwell fields in order to be consistent with the assumed homothety. Does it imply $\cL_\xi \Phi=-\kappa={\rm constant}$ and $\cL_\xi F_{\mu\nu} \propto F_{\mu\nu}$? If so, how is the proportionality constant related with $\kappa$?
Once we know how $\Phi$ and $F$ behave under the homothety, we can plug it into the equations of motion to investigate spherically symmetric CSS solutions, which is the subject of Section~\ref{sec:SphericalCSS}.

%%%%%%%%%%%%%%%%%%%%%%%%%%%%%%%%%%%%%%%%%%
\subsection{Action of the homothety on the dilaton and Maxwell fields\label{sec:HomoConds}}
%%%%%%%%%%%%%%%%%%%%%%%%%%%%%%%%%%%%%%%%%%

In this section we make heavy use of identities concerning HVFs, which are displayed in Appendix~\ref{sec:AppA}.

\bigskip
Applying the Lie derivative to Eq.~\eqref{eq:EinsteinScalar} and using identity~\eqref{eq:commutator1} we obtain
\be
(\nabla^\sigma \Phi) \nabla_\sigma \left( \cL_\xi \Phi \right) = -2\big(2V+V'\,\cL_\xi\Phi\big) \,.
\label{eq:result1}
\ee

On the other hand, acting on the dilaton equation with the Lie derivative, and using relations~\eqref{eq:commutator2}, \eqref{eq:commutator3} and \eqref{eq:commutator4}, one finds
\be
\nabla^2 \left( \cL_\xi \Phi \right) -  a^2 e^{-2a\Phi} F^2 \cL_\xi \Phi + a e^{-2a\Phi} F^{\mu\nu} \cL_\xi F_{\mu\nu} - a e^{-2a\Phi} F^2 - \big(2V'+V''\,\cL_\xi\Phi\big) = 0\,.
\label{eq:result2}
\ee
If we consider pure Einstein-dilaton gravity --- by setting $F_{\mu\nu}=0$ and $V(\Phi)=0$ --- we get $\nabla^2 \left( \cL_\xi \Phi \right)=0$, which combined with~\eqref{eq:result1} implies $\cL_\xi \Phi ={\rm constant}$, as can be shown by using Stokes' theorem.

Applying the Lie derivative to the Maxwell equation we get
\be
-2a\nabla^\mu \left( \cL_\xi \Phi \right) F_{\mu\nu} - 2a \nabla^\mu \Phi \, \cL_\xi F_{\mu\nu} + \nabla^\mu \left( \cL_\xi F_{\mu\nu} \right) = 0\,.
\label{eq:result3}
\ee
If we take the divergence of this equation we obtain a trivial identity.

Acting with the Lie derivative on the (traceless part of the) Einstein equation yields
\bea
\!\!\!\!
2\nabla_\mu \left( \cL_\xi \Phi \right) \nabla_\nu \Phi 
+ 2\nabla_\nu \left( \cL_\xi \Phi \right) \nabla_\mu \Phi 
+ 2g_{\mu\nu} \big(2V+V'\,\cL_\xi\Phi\big) \qquad\qquad\qquad\qquad\qquad\qquad \nn\\
\!\!\!\!
+ e^{-2a\Phi} \left[ 2\left( \cL_\xi F_{\mu\sigma} \right) {F_\nu}^\sigma + 2\left( \cL_\xi F_{\nu\sigma} \right) {F_\mu}^\sigma - g_{\mu\nu} F^{\rho\sigma} \left( \cL_\xi F_{\rho\sigma} \right) - 4F_{\mu\sigma}{F_\nu}^\sigma + g_{\mu\nu} F^2 \right] \qquad\; \nn\\
\!\!\!\!
- 2a e^{-2a\Phi} \left( \cL_\xi \Phi \right) \left( 2F_{\mu\sigma}{F_\nu}^\sigma - \frac{1}{2} g_{\mu\nu} F^2\right)=0\,.  \qquad\qquad\qquad\qquad\qquad\qquad\qquad\qquad\quad\;
\label{eq:result4}
\eea
We recover~\eqref{eq:result1} by taking the trace of this equation. If we instead contract with $F^{\mu\nu}$, all the terms vanish independently and we get a trivial identity.
However, if we take the divergence of~\eqref{eq:result4} (and use the previous results, as well as the Bianchi identity for $F_{\mu\nu}$) we get a simple constraint:
\be
F^{\rho\sigma} \Big[ \nabla_\rho \left(\cL_\xi F_{\nu\sigma}\right) + \nabla_\sigma \left(\cL_\xi F_{\rho\nu}\right) + \nabla_\nu \left(\cL_\xi F_{\sigma\rho}\right) \Big] =0\,.
\label{eq:result5}
\ee
This means that the Lie derivative of the electromagnetic field strength, $\cL_\xi F_{\mu\nu}$, also satisfies the Bianchi identity, at least when contracted with $F^{\mu\nu}$.

%%%%%%%%%%%%%%%%%%%%%%%%%%%%%%%%%%%%%%%%%%
\subsection{Solving the homothetic conditions\label{sec:SolveHomConds}}
%%%%%%%%%%%%%%%%%%%%%%%%%%%%%%%%%%%%%%%%%%

In order to solve the homothetic conditions obtained above, we assume that $\nabla_\mu\Phi$ is a timelike vector. Although this is a restriction on the collapse of scalar fields, which generally does not satisfy this condition throughout the whole evolution, it appears to be justified for the case of exactly continuously self-similar collapses\footnote{We have checked this explicitly for the solutions obtained in Section~\ref{sec:SphericalCSS}.}.

First, it is convenient to express~\eqref{eq:result4} in terms of the dilaton stress-energy~\eqref{eq:Tdil} and the following tensor:
\be
E_{\mu\nu} \equiv e^{2a\Phi}T^{\rm (EM)}_{\mu\nu} = \frac{1}{4\pi} \left(F_{\mu\sigma}{F_\nu}^\sigma-\frac{1}{4}g_{\mu\nu}F^2\right)\,.
\ee
The result becomes significantly more compact, namely
\be
\cL_\xi E_{\mu\nu} - 2a (\cL_\xi\Phi) E_{\mu\nu} = - e^{2a\Phi} \cL_\xi T_{\mu\nu}^{\rm (dil)}\,.
\label{eq:LieTEMLieTdil}
\ee

It can be shown that if $\nabla_\mu\Phi$ is a timelike vector, then the assumption of homothety implies that the left and right sides vanish independently [see Appendix~\ref{sec:AppB}].
This is not surprising, given that a similar result holds in the case of Einstein-Maxwell theory with a perfect fluid source~\cite{Wainwright:1976uu,Wainwright:1976ux}, and taking into account the equivalence between scalar fields and perfect fluids~\cite{Faraoni:2012hn}.
Moreover, it turns out that
\be
\cL_\xi V = -2V \,,  \qquad
\cL_\xi \nabla_\mu\Phi = 0\,.
\label{eq:LieVLieDPhi}
\ee
On account of Eq.~\eqref{eq:commutator1}, it follows immediately that
\be
\cL_\xi \Phi = -\kappa\,,
\label{eq:homo_dilaton2}
\ee
with $\kappa$ a constant. Therefore, condition~\eqref{eq:result1} is automatically satisfied.
Note that the identities~\eqref{eq:LieVLieDPhi} also imply
\be
\cL_\xi V' = -2V'\,.
\label{eq:LieDV}
\ee

To figure out the implications of
\be
\cL_\xi E_{\mu\nu} = 2a (\cL_\xi\Phi) E_{\mu\nu} 
\label{eq:LieEmunu}
\ee
for the action of the homothety on the Maxwell field we adapt the methods of~\cite{Wainwright:1976uu,Wainwright:1976ux}. The procedure depends on whether the field strength is null or not. Technically, a null electromagnetic field $F_{\mu\nu}$ is characterized by the conditions
\be
F_{\mu\nu}F^{\mu\nu}=0=F_{\mu\nu}\star\!F^{\mu\nu}\,.
\ee
We now analyze the two cases separately.

\paragraph{Non-null electromagnetic field.}
Assume first that the Maxwell field strength is non-null. In this case, it is a general result that $F_{\mu\nu}$ can be expressed in terms of the two distinct principal null directions, $k_\mu$ and $n_\mu$, which furthermore may be normalized so that $k^\mu n_\mu=-1$:
\be
F_{\mu\nu} = \tau_{\mu\nu} \cos\alpha + \star \tau_{\mu\nu} \sin\alpha\,,
\qquad
\textrm{where} \;\;
\tau_{\mu\nu} \equiv  \sqrt{8\pi} f \left( k_\mu n_\nu-n_\mu k_\nu \right)\,.
\label{eq:def_tau}
\ee
Here, $\alpha$ and $f$ are scalar quantities.
The electromagnetic stress-energy tensor can then be written in the form~\cite{Wainwright:1976uu}
\be
E_{\mu\nu} = f^2 \left[ g_{\mu\nu} + 2\left( k_\mu n_\nu + n_\mu k_\nu \right) \right]\,, 
\qquad
f^2 = \frac{1}{2}\sqrt{E_{\mu\nu}E^{\mu\nu}}\,.
\label{eq:def_Emunu}
\ee
It then follows that
\be
\cL_\xi E_{\mu\nu} = -2(1-a\cL_\xi \Phi) E_{\mu\nu} 
+ 2f^2 \left[ g_{\mu\nu} + k_\mu\cL_\xi n_\nu + n_\nu\cL_\xi k_\mu + k_\nu\cL_\xi n_\mu + n_\mu\cL_\xi k_\nu \right] \,.
\ee
When combined with Eq.~\eqref{eq:LieEmunu}, this implies
\be
k_\mu\cL_\xi n_\nu + n_\nu\cL_\xi k_\mu + k_\nu\cL_\xi n_\mu + n_\mu\cL_\xi k_\nu =
2\left(k_\mu n_\nu+k_\nu n_\mu\right)\,.
\ee
Contracting with $k^\mu$ and $n^\mu$ we find, respectively,
\begin{flalign}
\cL_\xi k_\nu &= \left( 2+k^\mu \cL_\xi n_\mu \right) k_\nu\,, \\
\cL_\xi n_\nu &= \left( 2+n^\mu \cL_\xi k_\mu \right) n_\nu\,.
\end{flalign}

Then, from the definition of $\tau_{\mu\nu}$ in~\eqref{eq:def_tau}, 
\be
\cL_\xi \tau_{\mu\nu} = \left(1+a \cL_\xi\Phi \right) \tau_{\mu\nu}\,,
\ee
and, using $\star \tau_{\mu\nu} \equiv \frac{1}{2}\epsilon_{\mu\nu\rho\sigma} \tau^{\rho\sigma}$ and Eq.~\eqref{eq:HKVF5},
\be
\cL_\xi \star\! \tau_{\mu\nu} = \left(1+a \cL_\xi\Phi \right) \star\! \tau_{\mu\nu}\,.
\ee
Therefore, we conclude that
\be
\cL_\xi F_{\mu\nu} =  \left(1+a \cL_\xi\Phi \right) F_{\mu\nu} + \widetilde{\kappa}\, \star\! F_{\mu\nu}\,,
\label{eq:LieFmunu}
\ee
where $\widetilde{\kappa}= \cL_\xi\alpha$.
$\blacksquare$

%\bigskip
\paragraph{Null electromagnetic field.}
In the null case the electromagnetic field strength can be written as~\cite{Wainwright:1976ux}
\be
F_{\mu\nu} = K_{\mu}A_{\nu}-K_{\nu}A_{\mu}\,,
\label{eq:nullcaseFmunu}
\ee
in terms of its repeated principal null direction $K^\mu$ and a spacelike vector $A^\mu$ that is orthogonal to $K^\mu$. Similarly, the Hodge dual can be expressed as
\be
\star F_{\mu\nu} = K_{\mu}B_{\nu}-K_{\nu}B_{\mu}\,,
\label{eq:nullcaseFstar}
\ee
where $B^\mu$ is orthogonal to both $K^\mu$ and $A^\mu$.
Then,
\be
E_{\mu\nu} = \frac{1}{4\pi}A_\sigma A^\sigma K_{\mu}K_{\nu}\,.
\label{eq:Emunu}
\ee
Acting with the Lie derivative on this expression, one finds
\be
\cL_\xi E_{\mu\nu} = -2E_{\mu\nu} + \frac{K_\mu K_\nu}{2\pi} A^\sigma \cL_\xi A_\sigma + \frac{A^\sigma A_\sigma}{4\pi} \left[ K_\mu \cL_\xi K_\nu + K_\nu \cL_\xi K_\mu \right]\,.
\label{eq:LieEmunuLONG}
\ee
Using~\eqref{eq:LieEmunu}, this becomes
\be
2K_\mu K_\nu A^\sigma \cL_\xi A_\sigma + A^\sigma A_\sigma \left[ K_\mu \cL_\xi K_\nu + K_\nu \cL_\xi K_\mu \right] = 8\pi \left( 1+a \cL_\xi \Phi \right) E_{\mu\nu}\,.
\label{eq:nullcase}
\ee

It is now convenient to introduce a second null vector $N^\mu$ such that
\be
N_\mu K^\mu =-1\,, \qquad
N_\mu A^\mu =0\,, \qquad
N_\mu B^\mu =0\,.
\ee
Contracting the previous equation with $N^\mu$ yields an expression for $\cL_\xi K_\nu$ that is proportional to $K_\nu$. However, one may always rescale the vectors $K^\mu\to\chi K^\mu$, $A^\mu\to\chi^{-1} A^\mu$, $B^\mu\to\chi^{-1} B^\mu$ so that
\be
\cL_\xi K_\nu = K_\nu\,.
\label{eq:nullcaseLieK}
\ee
Plugging this back in~\eqref{eq:nullcase} we get
\be
A^\sigma \cL_\xi A_\sigma = a (\cL_\xi \Phi) A^\sigma A_\sigma \,.
\ee
Note that orthogonality between $A^\mu$ and $K^\mu$ then implies
\be
K^\mu \cL_\xi A_\mu = -A^\mu \cL_\xi K_\mu = -A^\mu K_\mu = 0\,.
\ee
Therefore, the vector $\cL_\xi A_\mu - a(\cL_\xi \Phi) A_\mu$ lies in the orthogonal complement of $\textit{span}\{K^\mu,A^\mu\}$. It follows that
\be
\cL_\xi A_\mu - a(\cL_\xi \Phi) A_\mu = \theta K_\mu + \widetilde{\kappa} B_\mu\,,
\ee
for some scalars $\theta$ and $\widetilde{\kappa}$.
Inserting this result and~\eqref{eq:nullcaseLieK} back in Eq.~\eqref{eq:nullcaseFmunu}, taking also~\eqref{eq:nullcaseFstar} into account, we again obtain~\eqref{eq:LieFmunu}.
$\blacksquare$

\bigskip
Therefore, irrespectively of whether the electromagnetic field is null or non-null, the most general homothetic transformation laws for the dilaton and gauge fields are
\begin{subequations}
\bea
\cL_\xi\Phi &=& - \kappa\,, \\
\cL_\xi F_{\mu\nu} &=&  (1-a \kappa) F_{\mu\nu} +  \widetilde{\kappa} \star\! F_{\mu\nu} \,.
\eea
\label{eq:LiePhiLieF}
\end{subequations}
\!\!\!The only assumption employed in deriving this result was that the gradient of the dilaton is timelike.

Plugging results~\eqref{eq:LiePhiLieF} and~\eqref{eq:LieDV} back in~\eqref{eq:result2} one obtains the restriction
\be
\widetilde{\kappa}\, F^{\mu\nu}\star\! F_{\mu\nu} = 0\,.
\ee
Therefore, to be consistent with continuous self-similarity either $\widetilde{\kappa}=0$ or the invariant $F^{\mu\nu}\star\! F_{\mu\nu}$ must vanish, as is the case for a purely electric (or magnetic) Maxwell field\footnote{This statement assumes $a\neq0$, otherwise we just recover the Einstein-Maxwell system.}.

On the other hand, Eq.~\eqref{eq:result3}, together with~\eqref{eq:MaxwellEOM} and~\eqref{eq:LiePhiLieF}, yields
\be
(\nabla^\mu \widetilde{\kappa} - 2a \widetilde{\kappa} \nabla^\mu\Phi) \star\! F_{\mu\nu} =0\,,
\label{eq:tildek}
\ee
which can be seen as a condition on the scalar $\widetilde{\kappa}$.
A simple solution is given by
\be
\widetilde{\kappa} = \eta\, e^{2a\Phi}\,,
\ee
where $\eta$ is constant, but more generally one only needs the expression within parenthesis to be orthogonal to $\star F_{\mu\nu}$. This is satisfied in the case of spherically symmetric and purely electric configurations, for example.

\bigskip
Before we end this section, let us consider what kind of dilaton potential $V(\Phi)$ is consistent with continuous self-similarity of the spacetime. This is determined by solving condition~\eqref{eq:LieVLieDPhi}, supplemented by~\eqref{eq:homo_dilaton2}.
It then follows immediately that our results apply to the class of potentials of the Liouville type,
\be
V(\Phi)= \Lambda\, e^{\mu\Phi}\,,
\label{eq:LiouvillePotential}
\ee
as long as
\be
\kappa\,\mu = 2\,.
\label{eq:kappabeta}
\ee
This observation leads to an interesting result. In the presence of a Liouville potential, the constant $\kappa$ in the homothetic transformation of the dilaton (which is arbitrary when $V=0$) gets fixed in terms of the coefficient $\mu$ controlling the exponential behavior of the potential.

%%%%%%%%%%%%%%%%%%%%%%%%%%%%%%%%%%%%%%%%%%
\mysection{Spherically symmetric CSS solutions\label{sec:SphericalCSS}}
%%%%%%%%%%%%%%%%%%%%%%%%%%%%%%%%%%%%%%%%%%

In this section we investigate the effects of including an electromagnetic field (with a coupling to the dilaton) on the continuously self-similar spherical collapse of a scalar field.

The approach we take parallels that of Brady~\cite{Brady:1994aq}, with the difference that we include an additional Maxwell field that couples to the scalar. The idea is to cast the equations of motion as an autonomous system, the integral curves of which determine specific CSS solutions. The critical exponents can then be read off from the relevant modes of the linearized problem around the fixed points of the system.

One may also ask what happens if we include a Liouville potential consistent with continuous self-similarity. We derive the governing CSS equations for that case in Appendix~\ref{sec:AppC}, concluding also that it does not affect the critical exponent.

\bigskip
We adopt spherical symmetry from now on, and employ Bondi coordinates in which the line element reads
\be
 ds^2 = -g(u,r)\overline{g}(u,r)\, du^2 - 2g(u,r) du dr + r^2 \left[d\theta^2 + \sin^2\theta \,d\varphi^2 \right]\,.
\label{eq:metric_bondi}
\ee
Here $u$ is a retarded null coordinate.
Assuming continuous self-similarity, the metric can be put in the form
\be
 ds^2 = -g(x)\overline{g}(x)\, du^2 - 2g(x) du dr + r^2 d\Omega^2\,,
\label{eq:metric_bondiSS}
\ee
where $x\equiv r/|u|$ and $d\Omega^2$ represents the line element on the unit sphere, $d\theta^2 + \sin^2\theta \,d\varphi^2$. The homothetic vector field for such a metric is
\be
\xi = u \frac{\partial}{\partial u} + r \frac{\partial}{\partial r}\,.
\label{eq:HKVF}
\ee

Recall that the Maxwell field is compatible with continuous self-similarity only if it is purely electric or purely magnetic. Taking this into account it is not hard to show that, for spherically symmetric configurations,  condition~\eqref{eq:LieFmunu} actually implies $\widetilde{\kappa}=0$ .
Therefore, the most general homothetic transformations of the matter fields consistent with the equations of motion in spherical symmetry are
\be
\cL_\xi \Phi=-\kappa \,, \qquad \cL_\xi F_{\mu\nu}=(1-a\kappa)F_{\mu\nu}\,,
\ee
i.e., any additional contribution proportional to $\star F_{\mu\nu}$ in $\cL_\xi F_{\mu\nu}$ necessarily vanishes.

In the following we analyze the purely electric case in detail. The purely magnetic case is treated in Appendix~\ref{sec:AppD}, where it is shown to reduce to a formally identical autonomous system.

%%%%%%%%%%%%%%%%%%%%%%%%%%%%%%%%%%%%%%%%%%
\subsection{Reducing the equations of motion to an autonomous system}
%%%%%%%%%%%%%%%%%%%%%%%%%%%%%%%%%%%%%%%%%%

From now on we restrict to a purely electric Maxwell field.
In terms of the Bondi coordinates used to express the line element as in~\eqref{eq:metric_bondiSS}, the transformation of the matter fields under homothety implies
\be
\Phi=\phi(x)-\kappa \log(u/u_0) \,, \qquad F= -\frac{2q(x)}{|u|^{1+a \kappa}} du \wedge dr\,,
\label{eq:ansatzPhiF}
\ee
where $u_0$ is an arbitrary (null-)time scale. In the following we set it to $u_0=1$.

Expressed in terms of the self-similar variable $x$, the Einstein equations~\eqref{eq:EinsteinEOM} become
\bea
&& (x \overline{g})' = g (1-\omega^2)\,, \\
&& x g' = g \gamma^2\,, \\
&& g-\overline{g} = 2\kappa^2 - (\overline{g}-2x)(\gamma^2+2\kappa\gamma) + g \omega^2\,,
\eea
where the prime stands for a derivative with respect to variable $x$ and
\bea
&& \phi(x) \equiv \int^x_0 \frac{\gamma(\hat{x})}{\hat{x}}d\hat{x}\,, \\
&& \omega(x) \equiv \frac{x e^{-a\phi(x)} q(x)}{g(x)}\,.
\label{eq:defomega}
\eea
The Maxwell equations~\eqref{eq:MaxwellEOM} yield
\bea
&& x q g' + g \left[q(2a\gamma-2) - x q'\right] = 0\,, \\
&& x q g' + g \left[q(2a\gamma-1+a\kappa) - x q'\right] = 0\,,
\eea
from which it immediately follows that either $q(x)=0$ or\footnote{The metric function $g(x)$ should not vanish, otherwise the spacetime would be singular.}
\be
a\kappa = -1\,.
\label{eq:akappa}
\ee
The first option simply states that the Maxwell field vanishes so it recovers the Einstein-dilaton self-similar system studied in Ref.~\cite{Brady:1994aq}.
Therefore, from now on we will assume that the homothety parameter $\kappa$ and the dilaton coupling $a$ obey~\eqref{eq:akappa}.
Using Eq.~\eqref{eq:defomega} to replace $q(x)$ with $\omega(x)$, the only non-trivial Maxwell equation is expressed as
\be
x \omega' = -\omega\left(1+\frac{\gamma}{\kappa}\right)\,.
\ee

The dilaton equation~\eqref{eq:dilatonEOM} turns out to be
\be
x(\overline{g}-2x)\gamma' = 2\kappa x - \gamma(g-2x) + \left(\gamma-\frac{1}{\kappa}\right)g\omega^2\,,
\ee
and it is in fact implied by the Einstein and Maxwell equations as long as $\gamma(x)\neq -\kappa$.

Employing the field redefinitions
\be
y \equiv \frac{\overline{g}}{g}\,, \qquad
z \equiv \frac{x}{\overline{g}}\,,
\label{eq:defyz}
\ee
and implementing the change of coordinate
\be
\xi = \log x\,,
\ee
the field equations are converted into an autonomous system:
\begin{subequations}
\bea
\dot{y} &=& 1-\omega^2 - y(1+\gamma^2)\,, \label{eq:AUTOy}\\
\dot{z} &=& z\left( 2-\frac{1-\omega^2}{y} \right)\,, \label{eq:AUTOz}\\
\dot{\omega} &=& -\omega \left( 1+\frac{\gamma}{\kappa} \right)\,, \label{eq:AUTOomega}\\
(1-2z) \dot{\gamma} &=& 2\kappa z -\gamma \left(\frac{1}{y}-2z\right) + \frac{\omega^2}{y} \left( \gamma-\frac{1}{\kappa} \right)\,, \label{eq:AUTOgamma}\\
(\gamma+\kappa)^2  &=&  \frac{1+\kappa^2-\frac{1-\omega^2}{y}}{1-2z} \,, \label{eq:AUTOconstraint}
\eea
\label{eq:AUTO}
\end{subequations}
\!\!\!where the overdot stands for a derivative with respect to $\xi$.
These equations reduce to the autonomous system obtained by Brady~\cite{Brady:1994aq} when the Maxwell field vanishes, $\omega=0$. As in Ref~\cite{Brady:1994aq}, equation~\eqref{eq:AUTOconstraint} is an algebraic constraint, which means that one only needs to care about the evolution equations for the two degrees of freedom coming from the metric, $y$ and $z$, plus one degree of freedom accounting for the electric field, $\omega$. The addition of the electric field enlarges the phase space of CSS solutions, which now becomes three-dimensional.

Since Eq.~\eqref{eq:AUTOconstraint} is quadratic in $\gamma$ it has generically two solutions, the positive leaf and the negative leaf, depending on the choice of sign when taking the square root. We are only interested in real solutions. A given solution can change leaf only when it hits the surface $y=\frac{1-\omega^2}{1+\kappa^2}$. Nevertheless, the uncharged plane $\omega=0$ has opposite character (attractive or repulsive) on the two leafs, as we discuss in Section~\ref{sec:PhaseSpace}.

Note that the autonomous system~\eqref{eq:AUTO} enjoys a $\mathbb{Z}_2$ symmetry under which $\gamma$ and $\kappa$ flip sign simultaneously:
\be
\gamma \to -\gamma\,, \qquad \kappa \to - \kappa\,.
\label{eq:Z2symmetry}
\ee
This descends from the reflection symmetry of the original equations of motion~\eqref{eq:EOM}, under which the dilaton field $\Phi$ and the dilaton coupling $a$ also flip sign simultaneously. Therefore, there is no loss of generality in considering only non-negative values of $\kappa$, as we will do from now on.

Every integral curve of the dynamical system~\eqref{eq:AUTO} corresponds to a spherically symmetric CSS solution of the Einstein-Maxwell-dilaton theory. Flowing along any given integral curve (by increasing $\xi$) can be thought of as moving to larger radial coordinate $r$ at fixed retarded null time $u$, since $\xi=\log(r/|u|)$. Naturally, we are interested in selecting only those solutions that have a regular origin. In practice, this condition determines an `initial condition' for integral curves of relevance, to which we now turn.

%%%%%%%%%%%%%%%%%%%%%%%%%%%%%%%%%%%%%%%%%%
\subsection{Regularity conditions at the origin\label{sec:regularity}}
%%%%%%%%%%%%%%%%%%%%%%%%%%%%%%%%%%%%%%%%%%

In the Bondi coordinates adopted, the origin is defined by $r=0$. Again, we follow~\cite{Brady:1994aq} and fix
\be
g(u,0) = \overline{g}(u,0) = 1\,,
\label{eq:OriginCond1}
\ee
which is nothing but a choice of normalization of the coordinate $u$: it corresponds to the proper (null-)time of an observer sitting at the origin. In terms of the fields introduced in~\eqref{eq:defyz}, this is expressed as
\be
y(x=0)=1\,, \qquad
z(x=0)=0\,.
\label{eq:initcondyz}
\ee

A straightforward calculation reveals that the total stress-energy tensor, evaluated at the origin and using~\eqref{eq:OriginCond1}, behaves as
\be
8\pi T_{\mu\nu} \stackrel{r\to0}{\longrightarrow}
\left(
\begin{array}{cccc}
 \frac{\gamma(0)^2+\omega(0)^2}{r^2} & \frac{\gamma(0)^2+\omega(0)^2}{r^2} & 0 & 0 \\
 \frac{\gamma(0)^2+\omega(0)^2}{r^2} & \frac{2 \gamma(0)^2}{r^2} & 0 & 0 \\
 0 & 0 & \omega(0)^2-\gamma(0)^2 & 0 \\
 0 & 0 & 0 & \sin^2\theta \left(\omega(0)^2-\gamma(0)^2\right) \\
\end{array}
\right)\,.
\ee
Therefore, in order to have a non-singular center one must impose
\be
\gamma(x=0)=0\,, \qquad
\omega(x=0)=0\,.
\label{eq:initcondgo}
\ee
The same result follows from evaluating the Ricci and Kretschmann scalars, since these quantities are given respectively by
\be
R \stackrel{r\to0}{\longrightarrow} \frac{2 \gamma(0)^2}{r^2}\,, \qquad
K \stackrel{r\to0}{\longrightarrow} R_{\mu\nu\rho\sigma}R^{\mu\nu\rho\sigma}=\frac{4 \left( 2\gamma(0)^4 - 4\gamma(0)^2 \omega(0)^2 + 5\omega(0)^4 \right)}{r^4}\,.
\ee

Thus, our initial conditions are fully specified by Eqs.~\eqref{eq:initcondyz} and~\eqref{eq:initcondgo}. It is easily checked that these initial conditions are consistent with the constraint equation~\eqref{eq:AUTOconstraint}.

%%%%%%%%%%%%%%%%%%%%%%%%%%%%%%%%%%%%%%%%%%
\subsection{Local analysis of the autonomous system}
%%%%%%%%%%%%%%%%%%%%%%%%%%%%%%%%%%%%%%%%%%

Given the constraint~\eqref{eq:AUTOconstraint}, the autonomous system lives effectively in the three-dimensional phase space $\{y(\xi),z(\xi),\omega(\xi)\}$.
Clearly, real solutions can exist only if the right side of Eq.~\eqref{eq:AUTOconstraint} is non-negative. This selects two domains in the phase space where physical solutions can be supported:
\bea
z\leq\frac{1}{2}  & \text{and}  & y\geq\frac{1-\omega^2}{1+\kappa^2}  \label{eq:lowerdomain}\\
& \text{or} & \nn \\
z\geq\frac{1}{2}  & \text{and}  & y\leq\frac{1-\omega^2}{1+\kappa^2} \,. \label{eq:upperdomain}
\eea
Precisely at $z=1/2$ (and $y = \frac{1-\omega^2}{1+\kappa^2}$, as required by finiteness of the first derivative of the dilaton), the field $\gamma$ can be discontinuous, just like for the uncharged system studied by Brady~\cite{Brady:1994aq} ---and with similar consequences. This represents a line in the 3D phase space where the standard uniqueness theorem for systems of ordinary differential equations is not applicable. The upshot is that integral curves may have (depending on the value of $\kappa$, as we will see) a one-parameter family of possible continuations upon crossing this line.
Physically, this discontinuity line ---sometimes referred to as the self-similarity horizon--- represents a Cauchy horizon: solutions that reach this line require extra data to define how they are extended beyond it.

If one insists on $\gamma$ being continuous across this line, then Eq.~\eqref{eq:AUTOgamma} implies 
\be
\gamma = \frac{1}{\kappa} \left(1 - \frac{\omega^2}{\kappa^2 y}\right) = \frac{\kappa^2-\omega^2-2\kappa^2\omega^2}{\kappa^3(1-\omega^2)} \qquad \text{when} \quad z=\frac{1}{2}\,.
\ee
In fact, solutions with a regular origin %(whose integral curves start at $(y,z,\omega,\gamma)=(1,0,0,0)$) 
necessarily stick to the $\omega=0$ plane, as we discuss in Section~\ref{sec:PhaseSpace} below. Therefore, if such a solution reaches the discontinuity line it must do so at a crossing point located at
\be
{\bf C}:  \quad (y,z,\omega,\gamma) = \left( \frac{1}{1+\kappa^2},\frac{1}{2},0,\frac{1}{\kappa} \right)\,.
\ee
Note that even when there is infinite non-uniqueness of solutions past this crossing point, analyticity singles out only one spacetime, since it imposes the following directional derivatives for the integral curves at point {\bf C}:
\be
\frac{dz}{dy}\Big|_{\bf C} = -\frac{\kappa^2}{2}\,, \qquad\qquad
\frac{d\omega}{dy}\Big|_{\bf C} = 0\,.
\ee

\bigskip
Let us now analyze the fixed points of the autonomous system described by~\eqref{eq:AUTO}. In total there are five stationary points,
\be
\begin{array}{ll}
{\bf P1}:  & (y,z,\omega,\gamma) = \left( 1,0,0,0 \right)  \\
{\bf P2}^\pm:  & (y,z,\omega,\gamma) = \left( \frac{1}{2},\frac{1}{1\pm\kappa},0,\pm1 \right) \\
{\bf P3}^\pm:  & (y,z,\omega,\gamma) = \left( \frac{1}{\left(1+\kappa^2\right)^2},0,\pm\frac{\kappa}{\sqrt{1+\kappa^2}},-\kappa \right)\,.
\end{array}
\label{eq:fixedpoints}
\ee
Points ${\bf P1}$ and ${\bf P2}^\pm$ are exactly the same that appear in the self-similar collapse of a minimally coupled scalar field~\cite{Brady:1994aq}. In particular, ${\bf P1}$ is the `initial condition': solutions that are regular at the center are described by curves in the phase space that start at this point, when $x=0$ or equivalently when $\xi=-\infty$.

Points ${\bf P2}^\pm$ are saddle points: these are where the critical solutions end up (and also where near-critical solutions approach, before diverging along the repulsive direction). In particular, a local analysis around those points determines Choptuik's critical exponent. 

On the other hand, points ${\bf P3}^\pm$ are novel as they require a non-vanishing Maxwell field. They actually sit at the boundary of the lower domain in phase space~\eqref{eq:lowerdomain}. Note that in the limit $\kappa\to0$ they coincide with point ${\bf P1}$. However, this is not a limit we can take while working with the autonomous system~\eqref{eq:AUTO}. In fact, the eigenvalues obtained for the linearized system around ${\bf P3}^\pm$ do not match those from linearizing around ${\bf P1}$ when $\kappa\to0$ (see below). This apparent contradiction is resolved once we take into account that the condition $a\kappa = -1$ was imposed to derive the system of equations~\eqref{eq:AUTO} and that it cannot hold in this limit. In other words, we cannot have a non-trivial Maxwell field consistent with $\kappa = 0$ and hence the equations reduce to those of Ref.~\cite{Brady:1994aq}.

\bigskip
Next, we investigate the character of each of the stationary points.
Linearizing our autonomous system around ${\bf P1}$,
\be
(y,z,\omega,\gamma) = (1,0,0,0) + (\delta y, \delta z, \delta\omega, \delta\gamma)\,,
\ee
one obtains
\begin{subequations}
\bea
\dot{\delta y} &=& -\delta y\,, \\
\dot{\delta z} &=& \delta z\,, \\
\dot{\delta\omega} &=& -\delta\omega\,, \\
\dot{\delta\gamma} &=& 2\kappa \delta z -\delta\gamma\,.
\eea
\end{subequations}
This linear system has one positive eigenvalue ($+1$) and three degenerate negative eigenvalues ($-1$). The eigenvector associated with the positive eigenvalue is $(0,1,0,\kappa)$. So the `initial condition' point has a single growing mode,
\be
\delta z + \kappa \delta\gamma \sim e^\xi\,,
\ee
and all other modes decay.

\bigskip
Linearizing around points ${\bf P2}^\pm$,
\be
\big(y,z,\omega,\gamma \big) = \bigg(\frac{1}{2}, \frac{1}{1+\epsilon \kappa}, 0, \epsilon \bigg) + \big(\delta y, \delta z, \delta \omega, \delta \gamma \big)\,, 
\ee
where $\epsilon = \pm1$, we get the following linear system:
\begin{subequations}
\bea
\dot{\delta y} &=& -2\delta y - \epsilon \delta \gamma\,, \\
\dot{\delta z} &=& \frac{4}{1+\epsilon \kappa}\delta y\,, \\
\dot{\delta \omega} &=& -\left(1 + \frac{\epsilon}{\kappa}\right)\delta\omega\,, \\
\dot{\delta \gamma} &=& \frac{4\epsilon(\epsilon \kappa +1)}{\epsilon \kappa -1}\delta y + \frac{2(\kappa + \epsilon)(\epsilon \kappa +1)}{\epsilon \kappa -1}\delta z - \frac{2\epsilon \kappa}{\epsilon \kappa -1} \delta \gamma\,.
\eea
\end{subequations}

There are four distinct eigenvalues for this system:
\be
\lambda_{1,2} = \frac{-\epsilon \kappa \pm \sqrt{4 - 3\kappa^2}}{\epsilon \kappa -1}\,, \quad
\lambda_3 = -\left(1 + \frac{\epsilon}{\kappa}\right)\,, \quad
\lambda_4 = -2\,,
\label{eq:eigenvalues}
\ee
and the first three are different for the two fixed points ${\bf P2}^\pm$, since they depend on the value of $\epsilon$.
The first two eigenvalues precisely agree with Brady's results~\cite{Brady:1994aq}. For $0\leq\kappa^2<1$ both $\lambda_1$ and $\lambda_2$ are real and can be shown to have opposite signs. If $1<\kappa^2$ only one of the points ${\bf P2}^\pm$ is physically relevant (the other has negative $z$-coordinate). In that case, $\lambda_{1,2}$ are both real and negative if $1<\kappa^2\leq4/3$, or complex conjugate with negative real part if $\kappa^2>4/3$.
The eigenvalue associated to the contribution from the Maxwell field %(i.e., whose eigenvector lies along the $\omega$-direction)
is $\lambda_3$. Recalling we are restricting, without loss of generality, the parameter space to $\kappa\geq0$, we observe that this eigenvalue will \textit{always} be negative for ${\bf P2}^+$ and positive for ${\bf P2}^-$ (for the relevant regime $0\leq\kappa^2<1$).\footnote{One can consider $\kappa < 0$ by making use of the symmetry of the autonomous system~\eqref{eq:AUTO}, which sends $\kappa\to-\kappa$ and $\gamma\to-\gamma$. This mapping effectively makes $\gamma$ ``switch leaves'', as it is the solution of a quadratic equation~\eqref{eq:AUTOconstraint}. We see from Eq.~\eqref{eq:fixedpoints} that this interchanges ${\bf P2}^+$ and ${\bf P2}^-$, so that $\epsilon/\kappa>0$ now holds for ${\bf P2}^-$, which has a lower $z$-coordinate than ${\bf P2}^+$ when $\kappa$ is negative.}
Eigenvalue $\lambda_4$ comes from the constraint equation on $\gamma$ and is therefore redundant. 

Hence, for $\kappa^2>1$ only the fixed point ${\bf P2}^+$ matters and it is attractive (with spiralling character when $\kappa^2>4/3$). For $0\leq\kappa^2<1$, both ${\bf P2}^\pm$ are saddle points, but ${\bf P2}^+$ has a single growing mode, whereas ${\bf P2}^-$ has {\em two} relevant directions: besides the one found in Ref.~\cite{Brady:1994aq} for the uncharged case, there is another along the electric field direction. In a strict sense, ${\bf P2}^-$ is then not a critical point.

It is well-known that the critical exponent $\beta$ can be extracted from such local analysis around the critical point~\cite{Koike:1995jm}. As usual, it is obtained as the inverse of the real part of the eigenvalue associated with the growing mode,
\be
\beta = \frac{1}{\text{Re}\, \lambda_2} = \frac{1-\kappa}{\sqrt{4-3\kappa^2}-\kappa}\,.
\ee
When $\kappa=0$ ---for which there is an exact solution in closed form obtained by Roberts~\cite{Roberts:1989sk}--- the critical exponent simply reduces to $1/2$, but more generally we see it can take any value in the interval $[1/2,1/4)$ as $\kappa$ is varied between $0$ and $1$.

\bigskip
For completeness, we finally consider the linearization around the remaining stationary points, ${\bf P3}^\pm$. This yields the following linear system:
\begin{subequations}
\bea
\dot{\delta y} &=& -(1+\kappa^2)\delta y -\frac{2\epsilon \kappa}{\sqrt{1+\kappa^2}}\delta\omega +  \frac{2\kappa}{(1+\kappa^2)^2} \delta \gamma\,, \\
\dot{\delta z} &=& (1-\kappa^2)\delta z\,, \\
\dot{\delta \omega} &=& \frac{-\epsilon}{\sqrt{1+\kappa^2}}\delta\gamma\,, \\
\dot{\delta \gamma} &=& - 2\epsilon (1+\kappa^2)^{5/2}\delta\omega -(1+\kappa^2)\delta \gamma\,,
\eea
\end{subequations}
and the corresponding eigenvalues are
\be
\lambda_1 = 1-\kappa^2\,, \quad 
\lambda_2 = -(1+\kappa^2)\,, \quad 
\lambda_3 = 1+\kappa^2\,, \quad 
\lambda_4 = -2(1+\kappa^2)\,.
\ee
For $\kappa^2<1$ each of these points has two growing modes and two decaying modes. For $\kappa^2>1$ there is only one growing mode and three decaying modes.

%%%%%%%%%%%%%%%%%%%%%%%%%%%%%%%%%%%%%%%%%%
\subsection{Phase space of CSS solutions\label{sec:PhaseSpace}}
%%%%%%%%%%%%%%%%%%%%%%%%%%%%%%%%%%%%%%%%%%

\begin{figure}[!t]
\centering
\includegraphics[width=0.48\textwidth]{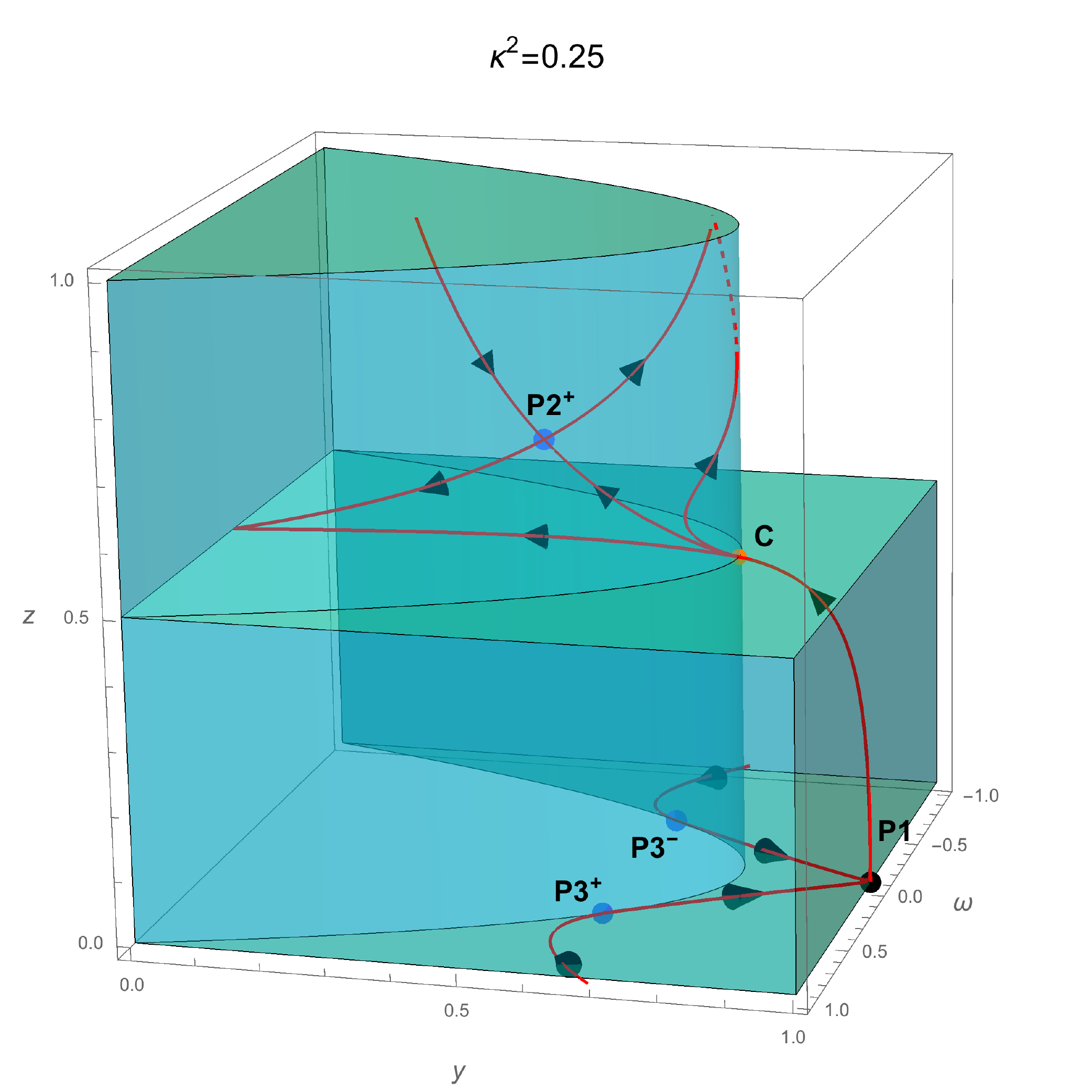}
%\quad
\includegraphics[width=0.48\textwidth]{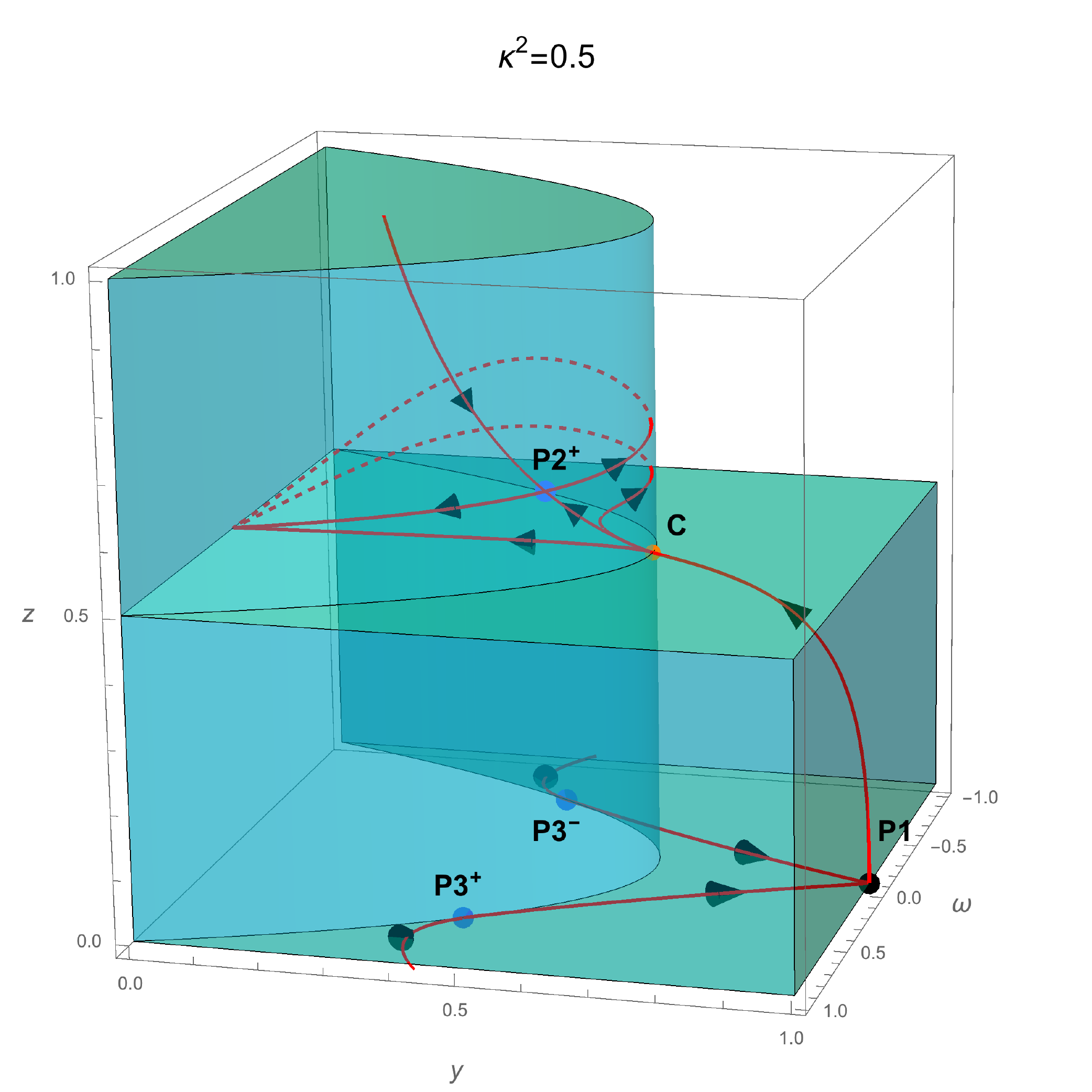}
\caption{Phase space of continuous self-similar (and spherically symmetric) solutions of Einstein-Maxwell-dilaton theory. The shaded regions indicate the two domains~\eqref{eq:lowerdomain} and~\eqref{eq:upperdomain} which can support real solutions. The left and right panels correspond to the choices $\kappa^2=0.25$ and $\kappa^2=0.5$, respectively. Solutions that are regular at the origin are represented by integral curves that start at point ${\bf P1}$. The critical solution is the one going through the crossing point ${\bf C}$ and ending at the fixed point ${\bf P2}^+$. In both panels the fixed point ${\bf P2}^-$ does not appear because its $z$-coordinate is outside of the range plotted. Other integral curves are included for illustrative purposes. Solid (dashed) curves indicate solutions that lie on the positive (negative) leaf of $\gamma$.}
\label{fig:phase_space1}
\end{figure}

\begin{figure}[!t]
\centering
\includegraphics[width=0.48\textwidth]{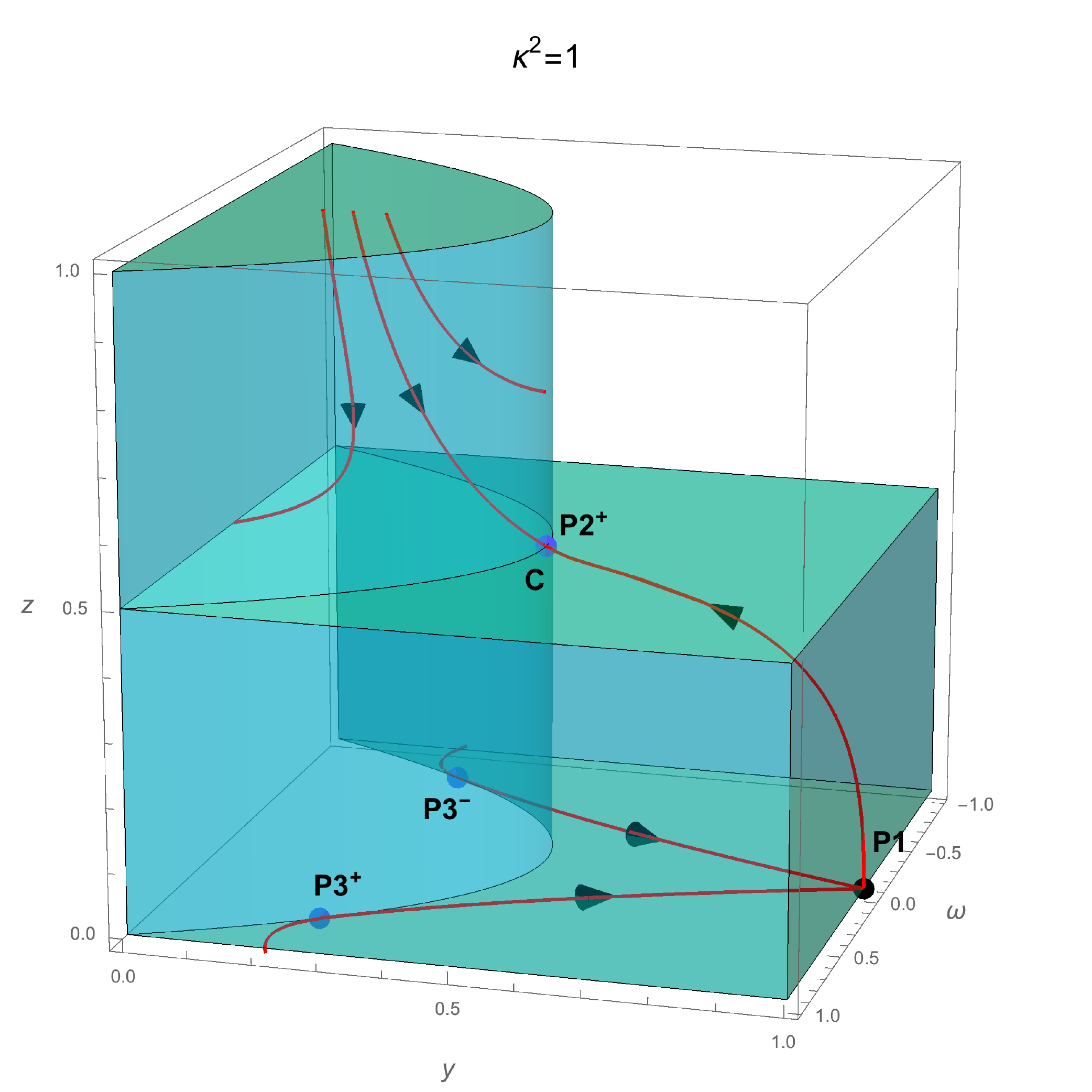}
%\quad
\includegraphics[width=0.48\textwidth]{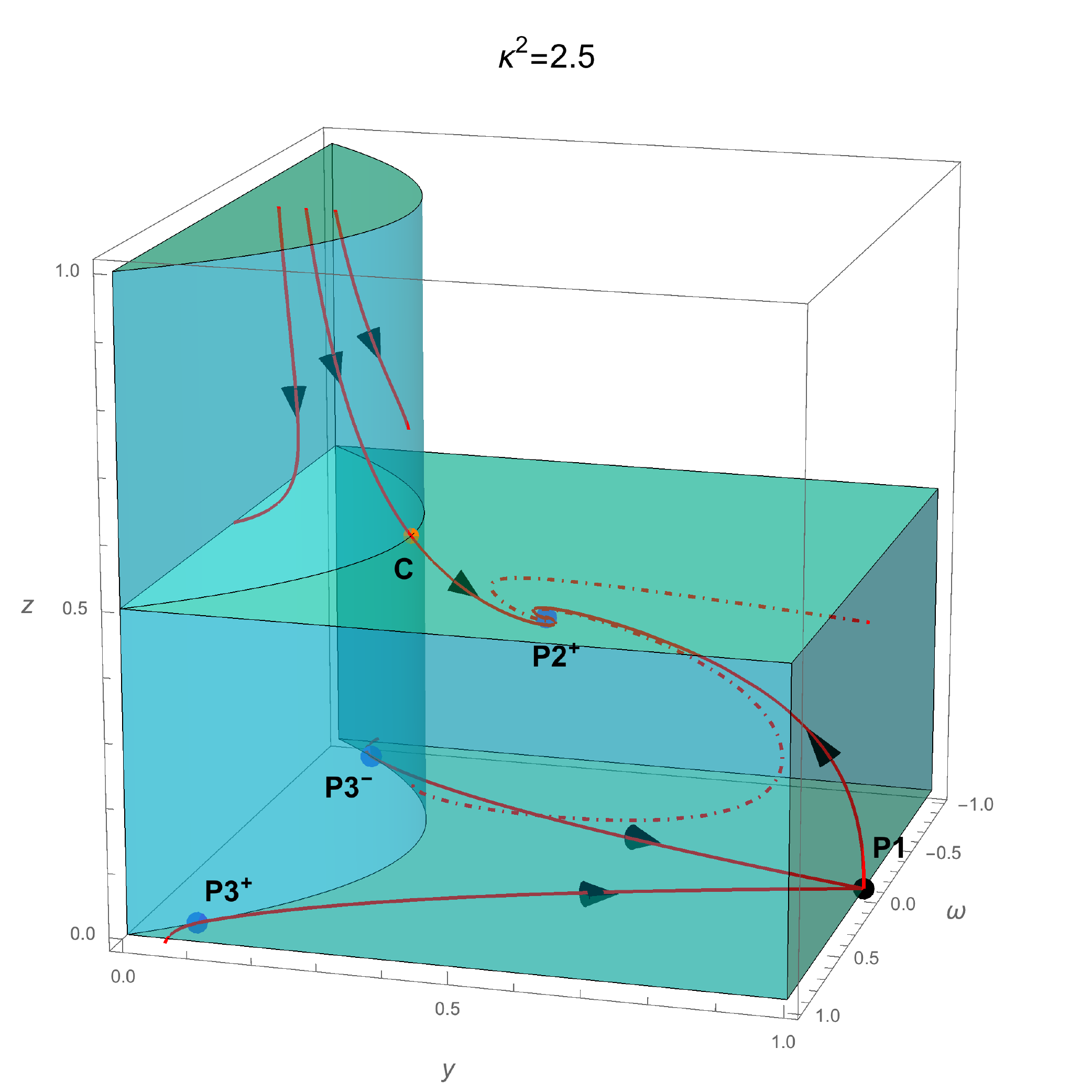}
\caption{Same as in Fig.~\ref{fig:phase_space1} but for larger values of $\kappa$. The left panel shows the phase space for $\kappa^2=1$. For this specific choice, and only in this case, the fixed point ${\bf P2}^+$ coincides with the would-be crossing point ${\bf C}$. This is the endpoint of the unique solution with a regular origin, i.e., that starts at ${\bf P1}$. The right panel displays the phase space for $\kappa^2=2.5$. The dot-dashed lines indicate other integral curves of the autonomous system corresponding to solutions without a regular center.}
\label{fig:phase_space2}
\end{figure}

We are now in condition to form a global picture of what the phase space of CSS solutions looks like. Shown in Fig.~\ref{fig:phase_space1} and Fig.~\ref{fig:phase_space2} are four examples of phase space solutions, differing in the value of $\kappa$. We grouped the figures in two pairs, since the qualitative behavior changes if we restrict ourselves to values $\kappa^2 < 1$ or to $\kappa^2 \geq 1$.
The curves plotted were obtained by numerically integrating Eqs.~\eqref{eq:AUTO}.

Consider first the behavior of integral curves in the case $\kappa^2 < 1$, presented in Fig.~\ref{fig:phase_space1}. According to Section~\ref{sec:regularity}, the curves that start from the point ${\bf P1}$ are the only regular solutions (at the origin). Following these solutions, one arrives at the crossing point ${\bf C}$, from which they can be continued in a non-unique fashion to the upper domain. However, there exists only one solution that ends up at the fixed point ${\bf P2}^+$ ---this is the critical solution. Nearby solutions are attracted to the critical curve for some time, but when the fixed point is approached they run off along the repulsive direction, forming either a black hole\footnote{Technically, what is formed is an apparent horizon.} (curves ending at $y=0$) or a naked singularity\footnote{As $z\to\infty$, $y$ approaches a constant value and a Cauchy horizon forms. It is not hard to show (see~\cite{Brady:1994aq}) that this surface is non-singular except at a point ($r=0=u$), which then corresponds to a naked singularity.} (curves for which $z\to\infty$). 

The fixed point ${\bf P2}^-$, despite being present for $\kappa^2 <1$, does not show up in either of the two panels of Fig.~\ref{fig:phase_space1} only because in both cases it is located above the plotted region. As mentioned before, ${\bf P2}^+$ is a critical point ---in the sense that it has only one relevant mode--- while ${\bf P2}^-$ is not: once we include the electric field it acquires a second relevant mode in the $\omega$-direction.

Taking into account Eqs.~\eqref{eq:AUTOomega} and~\eqref{eq:AUTOconstraint} it is straightforward to see that the `neutral' plane $\omega=0$ is an attractor for solutions in the positive leaf, but it is a repulsive surface for solutions in the negative leaf. We should note that solutions that are regular at the origin necessarily remain on the $\omega=0$ plane. This means that for regular solutions the electric field must vanish everywhere, so the addition of a Maxwell field is irrelevant. In particular, the electric field does not affect the critical exponent. One might think that the reason behind this result lies in the fact that $F^2$ is coupled to $e^{-2a\Phi}$ in the action~\eqref{eq:action} and, hence, its contribution would vanish if the dynamical system flows to $\Phi\to\infty$. However, this is not the case. Instead, this peculiarity is tied to $\omega=0$ being an attractor for the leaf in which the critical solution lies, and not so much to the coupling of the Maxwell field to the dilaton.

To illustrate more clearly what the phase space looks like, we also included a couple of integral curves going through fixed points ${\bf P3}^\pm$. These are the only curves shown that are not contained in the uncharged plane, $\omega=0$. They quickly approach point ${\bf P1}$ and then run along the curve connecting ${\bf P1}$ and ${\bf C}$, which is itself an attractor.

In the plots displayed in Fig.~\ref{fig:phase_space1}, a few curves are distinguished by a dashed linestyle, indicating those solutions live on the negative leaf of $\gamma$ [refer to Eq.~\eqref{eq:AUTOconstraint}]. A jump from the positive to the negative leaf occurs when an integral curve bounces off the surface $y=\frac{1-\omega^2}{1+\kappa^2}$ (and $z\neq1/2$). It is the continuation beyond this bounce in the upper domain that yields the only qualitative difference between the two panels. For lower $\kappa^2$ the critical solution shown marks the boundary between black hole formation and naked singularity formation. For larger $\kappa^2$ (but still less than unity) all solutions end up in black hole formation except for the critical solution itself.

We can now briefly analyze what happens in the cases $\kappa^2 \geq 1$, shown in Fig.~\ref{fig:phase_space2}. The left panel shows the phase space for $\kappa^2=1$, corresponding to the case in which the dilaton coupling takes the value $a=1$ dictated by heterotic string theory [see Eq.~\eqref{eq:akappa}]. Incidentally, it is only for this specific choice that the fixed point ${\bf P2}^+$ coincides with the would-be crossing point ${\bf C}$. In this case, this represents the endpoint of the unique solution with a regular origin, i.e., the one that starts at ${\bf P1}$. 
The right panel displays the phase space for $\kappa^2=2.5$. The only qualitative differences with respect to the left panel are that the fixed point ${\bf P2}^+$ is now in the interior of the lower domain and that the integral curves spiral in towards it. For $1<\kappa^2\leq4/3$, the fixed point ${\bf P2}^+$ acquires purely real (and negative) eigenvalues and so becomes an attractive point without spiral behavior. The dot-dashed lines indicate other integral curves of the autonomous system corresponding to solutions for which the origin is singular.

%%%%%%%%%%%%%%%%%%%%%%%%%%%%%%%%%%%%%%%%%%
\mysection{Discussion and outlook\label{sec:Conclusion}}
%%%%%%%%%%%%%%%%%%%%%%%%%%%%%%%%%%%%%%%%%%

In this work we presented the first study of continuously self-similar solutions of the source-free Einstein-Maxwell-dilaton system. Given the relevance of this model as a string-inspired alternative theory of gravity compatible with recent gravitational wave detections, it is worthwhile to pursue this avenue to attain an understanding of the theory's dynamics and implications.

We began by determining the conditions that the dilaton and Maxwell field must satisfy to be consistent with continuous self-similarity of the spacetime. Assuming only the gradient of the scalar field to be timelike, we were able to obtain the most general form of the homothetic transformations for the matter fields.

Then we examined spherically symmetric CSS solutions in the purely electric case. It was shown this problem can be understood in terms of a 3D dynamical system governed by autonomous equations, which greatly simplifies the task of mapping the phase space and determining the critical exponents. The upshot of this study is that the electric field becomes irrelevant at criticality and, in particular, it has no direct effect on the critical exponent. Nevertheless, the coupling between the gauge field and the scalar introduces some dependence of the critical exponent on the dilaton coupling constant.

It is also shown in the appendices that the inclusion of a Liouville potential or the consideration of a purely magnetic configuration does not spoil the autonomous property of the differential equations, so the same methods can be used. Once again, their contribution becomes irrelevant at criticality.

These results are in accordance with the statement that universality in critical collapse generally {\em cannot} be extended to mean the critical solution is universal among classes of matter models~\cite{Maison:1995cc}, in addition to independence on details of the initial conditions within a given theory. 

Our results on the homothetic conditions indicate, as a natural extension, that it would be interesting to augment the system with the axion field. This is strongly suggested by the appearance of terms proportional to $F^{\mu\nu}\star\! F_{\mu\nu}$ in Eq.~\eqref{eq:result2}, when $\widetilde{\kappa}\neq0$. Recall such terms are sources for the axion. It is conceivable that there exist homothetic actions in the Einstein-Maxwell-axion-dilaton system for which the dilaton and the axion mix. In fact, references~\cite{Hamade:1995jx,Eardley:1995ns} provide a clear indication that this is the case.

Another obvious generalization is the consideration of higher dimensions, which should be straightforward. In fact, it should be even simpler than in four dimensions, because only in the latter case can the Hodge dual of the field strength appear in the homothetic transformation law~\eqref{eq:LieFmunu}.

Yet another possible extension of interest, still in the context of low-energy string theories, is the consideration of multiple and/or non-abelian gauge fields.

In this paper we contemplated source-free solutions of the EMD equations of motion. We found that regular initial data implied the vanishing of the electric field throughout the entire continuous self-similar evolution. Nevertheless, solutions analogous to the Vaidya spacetime, sourced by charged null fluids, can be envisaged~\cite{Aniceto:2017gtx,Aniceto:2015klq}, thus allowing for a non-trivial Maxwell field. Within this arena, continuous self-similar solutions can be constructed analytically and we expect to report on this in the near future.

Finally, we have made heavy use of {\em continuous} self-similarity. That this kind of symmetry is the one that emerges at the threshold of black hole formation ---instead of discrete self-similarity--- is suggested by the work of~\cite{Hamade:1995jx}, but not proven. Therefore, a pertinent (and more challenging) problem that remains open concerns relaxing the assumption of continuous self-similarity to discrete self-similarity.

%%%%%%%%%%%%%%%%%%%%%%%%%%%%%%%%%%%%%%%%%%
\section*{Acknowledgments}
%%%%%%%%%%%%%%%%%%%%%%%%%%%%%%%%%%%%%%%%%%

It is our pleasure to thank Francesco Bigazzi, Roberto Emparan and Cristiano Germani for useful discussions. We are also grateful to Roberto Emparan for comments on a draft of this paper. 
We acknowledge financial support from the European Union's Horizon 2020 research and innovation programme under ERC Advanced Grant GravBHs-692951.
Funding for this work was partially provided by the Spanish MINECO under projects FPA-2016-76005-C2-2-P and MDM-2014-0369 of ICCUB (Unidad de Excelencia `Mar\'ia de Maeztu').

\bigskip

\appendix

%%%%%%%%%%%%%%%%%%%%%%%%%%%%%%%%%%%%%%%%%%
\mysection{Homothetic relations\label{sec:AppA}}
%%%%%%%%%%%%%%%%%%%%%%%%%%%%%%%%%%%%%%%%%%

In this appendix we collect useful identities concerning homothetic vector fields (HVFs) in four spacetime dimensions. From the definition
\be
\cL_\xi g_{\mu\nu} = 2g_{\mu\nu}
\qquad \Longleftrightarrow \qquad
\nabla_\mu \xi_\nu + \nabla_\nu \xi_\mu = 2g_{\mu\nu}\,,
\label{eq:HKVF1}
\ee
it follows that
\be
\nabla_\mu \xi^\mu = 4 \;\; \text{(in 4 dimensions)}\,,
\qquad
\nabla^2 \xi_\mu = -R_{\sigma\mu}\xi^\sigma\,,
\label{eq:HKVF2}
\ee
\be
\cL_\xi g^{\mu\nu} = -2g^{\mu\nu}\,,
\qquad
\cL_\xi {g^\mu}_\nu = 0\,.
\label{eq:HKVF3}
\ee

This has straightforward implications for the Riemann tensor and its contractions,
\be
\cL_\xi {R^\mu}_{\nu\rho\sigma} = 0\,,
\qquad
\cL_\xi R_{\mu\nu} = 0\,,
\qquad
\cL_\xi R = -2R\,,
\qquad
\cL_\xi G_{\mu\nu} = 0\,,
\label{eq:HKVF4}
\ee
where $G_{\mu\nu}=R_{\mu\nu}-\frac{1}{2}R g_{\mu\nu}$ is the Einstein tensor.
Eq.~\eqref{eq:HKVF1} also implies
\be
\nabla^\nu \xi_\sigma \nabla^\sigma \Phi \nabla_\nu \Phi = (\nabla \Phi)^2\,.
\ee
The action of the homothety on the Levi-Civita tensor is
\be
\cL_\xi \epsilon_{\mu\nu\rho\sigma} = 4 \epsilon_{\mu\nu\rho\sigma}\,.
\label{eq:HKVF5}
\ee

In Section~\ref{sec:SSconditions} we make heavy use of expressions involving commutators of covariant derivatives with the Lie derivative along the HVF. Concerning this, we note the following identities for scalar fields:
\bea
&& [\nabla_\sigma, \cL_\xi] \Phi = 0\,, \label{eq:commutator1}\\
&& [\nabla^2, \cL_\xi] \Phi = 2 \nabla^2 \Phi\,. \label{eq:commutator2}
\eea
The general expressions for the commutator of the covariant derivative with a Lie derivative applied to vectors and antisymmetric tensors are
\bea
&& [\nabla_\mu, \cL_X] A^\mu = -A^\lambda \nabla_\lambda \nabla_\mu X^\mu\,,\\
&& [\nabla_\mu, \cL_X] T^{\mu\nu} = -T^{\lambda\nu} \nabla_\lambda \nabla_\mu X^\mu
\quad \text{for } T^{\lambda\nu} \;\; \text{antisymetric.}
\eea
These simplify in the case of HVFs, $X=\xi$:
\bea
&& [\nabla_\mu, \cL_\xi] A^\mu = 0\,, \label{eq:commutator3}\\
&& [\nabla_\mu, \cL_\xi] T^{\mu\nu} = 0\,. \label{eq:commutator4}
\eea
%

%%%%%%%%%%%%%%%%%%%%%%%%%%%%%%%%%%%%%%%%%%
\mysection{Proof of $\cL_\xi T_{\mu\nu}^{\rm (EM)}=0=\cL_\xi T_{\mu\nu}^{\rm (dil)}$\label{sec:AppB}}
%%%%%%%%%%%%%%%%%%%%%%%%%%%%%%%%%%%%%%%%%%

Here we show that the left and right sides of Eq.~\eqref{eq:LieTEMLieTdil} must vanish independently, if $\nabla_\mu\Phi$ is timelike. As a bonus, the identities~\eqref{eq:LieVLieDPhi} will also emerge from this analysis.
As in Section~\ref{sec:SolveHomConds}, the cases in which the Maxwell field is null or non-null have to be dealt with separately.

\paragraph{Non-null electromagnetic field.}
For this case the proof proceeds along the lines of Ref.~\cite{Wainwright:1976uu}, where more details can be found.

The idea is to determine the eigenvectors $v_\mu^{(i)}$ of the total stress-energy tensor, together with their associated eigenvalues $\sigma^{(i)}$, and use
\be
-2\sigma^{(i)} v_\mu^{(i)} + {T_\mu}^\nu \cL_\xi v_\nu^{(i)}= \cL_\xi \left[ {T_\mu}^\nu v_\nu^{(i)} \right] = \cL_\xi \left[ \sigma^{(i)} v_\mu^{(i)} \right] = (\cL_\xi \sigma^{(i)}) v_\mu^{(i)} + \sigma^{(i)} \cL_\xi v_\mu^{(i)}\,.
\label{eq:AppB1}
\ee
If the eigenvector $v^{(i)}$ is not null, then contraction with $v^{(i)\mu}$ immediately produces a relation for the associated eigenvalue,
\be
\cL_\xi \sigma^{(i)} = -2 \sigma^{(i)}\,.
\label{eq:LieSigma}
\ee

Now, the eigenvalues depend on whether or not the timelike vector $\nabla_\mu\Phi$ is an eigenvector of $T_{\mu\nu}^{\rm (EM)}$.
Assuming it is, it can be easily shown that it must be expressed as
\be
\nabla_\mu\Phi = - k_\mu (n^\nu \nabla_\nu\Phi) - n_\mu (k^\nu \nabla_\nu\Phi) \equiv v_\mu^{(1)}\,,
\ee
where the null vectors $k^\mu$ and $n^\mu$ were introduced in Section~\ref{sec:SolveHomConds}. We can also define a spacelike vector
\be
v_\mu^{(2)} \equiv - k_\mu (n^\nu \nabla_\nu\Phi) + n_\mu (k^\nu \nabla_\nu\Phi)\,,
\ee
which is orthogonal to $v^{(1)}$, plus two other linearly independent spacelike vectors $v^{(3)}$ and $v^{(4)}$, orthogonal to both $k$ and $n$. The four vectors $v^{(i)}$ are all (non-null) eigenvectors of the total stress-energy tensor, and their associated eigenvalues are
\begin{subequations}
\bea
\sigma^{(1)} &=& -f^2 e^{-2a\Phi} + (\nabla\Phi)^2 - 2V\,,\\
\sigma^{(2)} &=& -f^2 e^{-2a\Phi} - (\nabla\Phi)^2 - 2V\,,\\
\sigma^{(3)} = \sigma^{(4)} &=& +f^2 e^{-2a\Phi} - (\nabla\Phi)^2 - 2V\,.
\eea
\end{subequations}
By applying identity~\eqref{eq:LieSigma} on judiciously chosen linear combinations of these eigenvalues we find
\be
\cL_\xi V = -2V \qquad \text{and}\qquad \cL_\xi\left[ (\nabla\Phi)^2 \right] = -2(\nabla\Phi)^2\,.
\label{eq:AppB6}
\ee
Given that the eigenspace to which $\nabla_\mu\Phi=v_\mu^{(1)}$ belongs is one-dimensional, it follows from~\eqref{eq:AppB1} that
\be
\cL_\xi \nabla_\mu\Phi = \chi \nabla_\mu\Phi\,,
\label{eq:LieDPhi1}
\ee
for some scalar $\chi$. When combined with~\eqref{eq:AppB6}, this implies that $\chi$ must vanish.
This completes the proof of identities~\eqref{eq:LieVLieDPhi}, from which it immediately follows that
\be
\cL_\xi T_{\mu\nu}^{\rm (dil)} = 0 \qquad \text{and} \qquad \cL_\xi T_{\mu\nu}^{\rm (EM)}=0\,.
\quad\blacksquare
\label{eq:AppB8}
\ee

The remaining case to be analyzed occurs when $\nabla_\mu\Phi$ is not an eigenvector of $T_{\mu\nu}^{\rm (EM)}$.
In this case, the vector $\hat{v}^{(2)}\equiv v^{(2)}$ is still an eigenvector of the total stress-energy tensor, but now $\nabla_\mu\Phi$ cannot be contained in the span of $k_\mu$ and $n_\mu$. Instead, its place is taken by the vector $\hat{v}^{(1)}_\mu$ defined (up to scale) as being orthogonal to both $\nabla_\mu\Phi$ and the principal null directions,
\be
\hat{v}^{(1)}_\mu \nabla^\mu\Phi = \hat{v}^{(1)}_\mu k^\mu = \hat{v}^{(1)}_\mu n^\mu = 0\,.
\ee
It is easy to check that $\hat{v}^{(1)}$ so defined is an eigenvector of the total stress-energy tensor.
With a little more effort one can obtain two more independent eigenvectors, which take the form
\be
\hat{v}_\mu^{(\pm)} = -k_\mu(n^\sigma \nabla_\sigma\Phi) -n_\mu(k^\sigma \nabla_\sigma\Phi) + \left(-\frac{e^{2a\Phi}}{2f^2}(\nabla\Phi)^2 - \frac{1}{2} \pm \frac{h\, e^{2a\Phi}}{4f^4}\right) \nabla_\mu\Phi\,,
\label{eq:AppB10}
\ee
where
\be
h^2 \equiv 4(\nabla\Phi)^4 + 4f^4 e^{-4a\Phi} + 8 e^{-2a\Phi} b^2  \qquad \text{and} \qquad
b^2 \equiv E_{\mu\nu} \nabla^\mu\Phi \nabla^\nu\Phi\,.
\ee
The associated eigenvalues are
\begin{subequations}
\bea
\hat{\sigma}^{(1)} &=& -f^2 e^{-2a\Phi} - (\nabla\Phi)^2 - 2V\,,\\
\hat{\sigma}^{(2)} &=& +f^2 e^{-2a\Phi} - (\nabla\Phi)^2 - 2V\,,\\
\hat{\sigma}^{(\pm)} &=& \pm\frac{h}{2} - 2V\,.
\eea
\end{subequations}
As before, applying identity~\eqref{eq:LieSigma} on linear combinations of these eigenvalues we again obtain~\eqref{eq:AppB6}. The rest follows as in the previous case, since Eqs.~\eqref{eq:AppB1} and~\eqref{eq:AppB10} allow to show that~\eqref{eq:LieDPhi1} holds also in this case. $\blacksquare$

\paragraph{Null electromagnetic field.}
When the Maxwell field is null, the procedure above does not work because the total stress-energy tensor does not have any real eigenvectors. Therefore, a different route must be taken.

The electromagnetic part of the stress-energy tensor can now be written as [refer to Section~\ref{sec:SolveHomConds}]
\be
T_{\mu\nu}^{\rm (EM)} = e^{-2a\Phi} E_{\mu\nu} = \frac{e^{-2a\Phi}}{4\pi}A_\sigma A^\sigma K_{\mu}K_{\nu}\,,
\ee
from which one computes its Lie derivative by using~\eqref{eq:LieEmunuLONG}. Contracting the resulting expression with various combinations of the vectors $K^\mu$, $A^\mu$ and $B^\mu$, and using Eq.~\eqref{eq:LieTEMLieTdil}, it can be shown that
\be
K^\mu \cL_\xi \nabla_\mu\Phi = A^\mu \cL_\xi \nabla_\mu\Phi = B^\mu \cL_\xi \nabla_\mu\Phi = 0\,.
\ee
In order to derive this, it is necessary that $K^\mu\nabla_\mu\Phi\neq0$. This automatically holds when $\nabla^\mu\Phi$ is timelike, since $K^\mu$ is a null vector. Given that $\cL_\xi \nabla_\mu\Phi$ is orthogonal to all three vectors $K^\mu$, $A^\mu$ and $B^\mu$, it must be proportional to $K^\mu$,
\be
\cL_\xi \nabla_\mu\Phi = \widetilde{\chi} K_\mu\,.
\label{eq:LieDPhi2}
\ee
We may now plug this expression in the left side of the following equality:
\be
N^\mu \left( \cL_\xi T_{\mu\nu}^{\rm (dil)} \right) K^\nu = - N^\mu \left( \cL_\xi T_{\mu\nu}^{\rm (EM)} \right) K^\nu = 0\,,
\ee
from which it follows that
\be
\cL_\xi V = -2V\,.
\label{eq:LieV}
\ee

Now, consider the trace of the stress-energy tensor and how it transforms under homothety.
Applying the Lie derivative along the homothetic vector gives
\be
\cL_\xi \left[ g^{\mu\nu} T_{\mu\nu}^{\rm (dil)} \right] = -2 \cL_\xi \left[ (\nabla\Phi)^2 \right] - 8 \cL_\xi V\,.
\ee
On the other hand, since $ \cL_\xi \left(T_{\mu\nu}^{\rm (dil)}+T_{\mu\nu}^{\rm (EM)}\right) \propto \cL_\xi G_{\mu\nu}=0$, we have
\be
\cL_\xi \left[ g^{\mu\nu} T_{\mu\nu}^{\rm (dil)} \right]
= -2g^{\mu\nu} T_{\mu\nu}^{\rm (dil)} - g^{\mu\nu} \cL_\xi T_{\mu\nu}^{\rm (EM)}
= 4(\nabla\Phi)^2 + 16 V\,.
\ee
Equating these two expressions and using~\eqref{eq:LieV} yields
\be
\nabla^\mu\Phi\, \cL_\xi \nabla_\mu\Phi=0\,.
\ee
Finally, plugging in Eq.~\eqref{eq:LieDPhi2} and recalling that $K^\mu\nabla_\mu\Phi\neq0$ we conclude that $\widetilde{\chi}$ must vanish. Once again, this completes the proof of identities~\eqref{eq:LieVLieDPhi}, and~\eqref{eq:AppB8} immediately follows. $\blacksquare$

%%%%%%%%%%%%%%%%%%%%%%%%%%%%%%%%%%%%%%%%%%
\mysection{CSS equations with a Liouville potential\label{sec:AppC}}
%%%%%%%%%%%%%%%%%%%%%%%%%%%%%%%%%%%%%%%%%%

Here we briefly consider the effects of including a Liouville potential $V(\Phi)$ in the CSS equations. Recall that consistency with the homothetic conditions imposes $\cL_\xi V=-2V$, or equivalently, $V(\Phi)=\Lambda e^{2\Phi/\kappa}$. This will add some terms to our autonomous system.
Continuous self-similar collapses in Einstein-dilaton gravity with exponential potentials, but with no gauge fields, were previously considered in Ref.~\cite{Zhang:2015rsa}.

We start with an action given by~\eqref{eq:action} and we solve the full system of equations of motion~\eqref{eq:EOM} under the assumption of a purely electric Maxwell field, Eq.~\eqref{eq:ansatzPhiF}. The resulting equations differ from those obtained without the Liouville potential only in the angular part of the Einstein equations and in the scalar field equation. Proceeding with the same steps as for the purely electric case, we get the following set of equations:
\begin{subequations}
\bea
\dot{y} &=& \left(1 - \omega^2\right) - y\left(1 + \gamma^2\right) - 2\Lambda e^{2(\phi/\kappa+\xi)}\,, \\
\dot{z} &=& z\left( 2-\frac{1-\omega^2}{y} + \frac{2\Lambda}{y}e^{2(\phi/\kappa+\xi)}\right)\,, \\
\dot{\omega} &=& -\omega\left(1 + \frac{\gamma}{\kappa}\right)\,, \\
\dot{\phi} &=& \gamma\,, \\
(1 - 2z)\dot{\gamma} &=& 2z\left(\kappa + \gamma\right) - \frac{1}{\kappa y}\left(\gamma\kappa\left(1 - \omega^2\right) + \omega^2 - 2\Lambda e^{2(\phi/\kappa+\xi)}\left(1 + \kappa\gamma\right)\right)\,, \\
(\gamma+\kappa)^2 &=& \frac{\left(1 +\kappa^2\right)y - 1 + \omega^2 + 2\Lambda e^{2(\phi/\kappa+\xi)}}{y\left(1 - 2z\right)} \,.
\eea
\end{subequations}
One can see that when the Liouville potential is included, $\phi$ appears explicitly, instead of just $\dot{\phi}$, so the phase space is enlarged. More importantly, now the dependence on $\xi$ is explicit, so in this form the system is no longer autonomous.
Fortunately, by formally introducing a new field, $\zeta(\xi) \equiv e^{\phi(\xi)/\kappa+\xi}$, we recover an autonomous system, so the methods used in Section~\ref{sec:SphericalCSS} can be equally applied here.

As before, the equation for $\dot{\gamma}$ is redundant, since it follows from the rest of equations under the condition that $\gamma \neq -\kappa$. However, the phase space is still too large to be manageable. We can reduce it to a 3-dimensional phase space by considering a vanishing electric field\footnote{We could of course set $\Lambda = 0$, which would reduce the system to the purely electric one. However, the goal here is to see the effects of the Liouville potential.}.
Hence, for $\omega = 0$ we are left with a simpler set of equations (we are also dropping the equation for $\dot{\gamma}$):
\begin{subequations}
\bea
\dot{y} &=& 1 - y\left(1 + \gamma^2\right) - 2\Lambda \zeta^2\,, \\
\dot{z} &=& z\left( 2-\frac{1-2\Lambda\zeta^2}{y} \right)\,, \\
\dot{\zeta} &=& \zeta \left( 1+\frac{\gamma}{\kappa} \right)\,, \label{eq:zetadot}\\
(\gamma+\kappa)^2 &=& \frac{\left(1 +\kappa^2\right)y - 1 + 2\Lambda \zeta^2}{y\left(1 - 2z\right)} \,.
\eea
\end{subequations}
These equations match those obtained in~\cite{Zhang:2015rsa}.

The fixed points of this system are easy to find. The points equivalent to ${\bf P1}$ and ${\bf P2}^\pm$ in the electrically charged case are also stationary points in the Liouville potential case, namely $(y,z,\zeta,\gamma)=(1,0,0,0)$ and $(y,z,\zeta,\gamma)=(1/2,(1\pm\kappa)^{-1},0,\pm1)$. Through an analysis identical to that of Section~\ref{sec:regularity}, it can be shown that solutions with regular origins must obey $z(r=0)=\gamma(r=0)=\zeta(r=0)=0$, and therefore `start' at the first these points. In addition, there is one further fixed point at 
\be
(y,z,\zeta,\gamma)=\left(\frac{1}{1-\kappa^4},0,\frac{|\kappa|}{\sqrt{-2\Lambda(1-\kappa^2)}},-\kappa\right)\,,
\ee
but only for $0\leq\kappa^2<1$ and in the case of a negative $\Lambda$.

Since $\zeta$ enters all equations only through squared powers, except for~\eqref{eq:zetadot}, the linearization of the system around fixed points will leave no trace of $\Lambda$, and therefore of the Liouville potential. This indicates the inclusion of the Liouville potential also cannot change the critical exponent.
In this respect, the self-similar collapse of a spherically symmetric scalar field with a Liouville potential is similar to what we obtained for the case of Einstein-Maxwell-dilaton theory. There is, however, one qualitative difference. The fixed point $(y,z,\zeta,\gamma)=(1/2,(1+\kappa)^{-1},0,1)$, which is the analogue of ${\bf P2}^+$, is no longer a critical point, in the sense that it has {\em two} relevant directions: in addition to the previously existing mode with eigenvalue $\lambda_2$ [see Eq.~\eqref{eq:eigenvalues}] there is one more growing mode coming from~\eqref{eq:zetadot}.

%%%%%%%%%%%%%%%%%%%%%%%%%%%%%%%%%%%%%%%%%%
\mysection{CSS equations for a purely magnetic Maxwell field\label{sec:AppD}}
%%%%%%%%%%%%%%%%%%%%%%%%%%%%%%%%%%%%%%%%%%

A purely magnetic Maxwell field yielding a spherically symmetric configuration\footnote{The Maxwell field is, strictly speaking {\em not} spherically symmetric, though the metric and scalar field are.} is given, in the coordinate system~\eqref{eq:metric_bondiSS}, by
\be
F = 2P \sin\theta\, d\theta \wedge d\varphi\,.
\ee
$P$ measures the magnetic charge and is necessarily constant in order to satisfy the Bianchi identity.
Such an electromagnetic field is invariant under the homothetic vector~\eqref{eq:HKVF} ---i.e., it satisfies $\cL_\xi F_{\mu\nu}=0$--- so one must have $a\kappa=1$ in this case.

Following the same steps as for the purely electric field, the equations of motion reduce to
\begin{subequations}
\bea
\dot{y} &=& 1 - y(1+\gamma^2) - P^2 e^{-2(\phi/\kappa+\xi)}\,, \\
\dot{z} &=& z\left( 2 - \frac{1}{y} + \frac{P^2}{y} e^{-2(\phi/\kappa+\xi)} \right)\,, \\
\dot{\phi} &=& \gamma\,, \label{eq:CSSwithPh} \\
(\gamma+\kappa)^2 &=& \frac{ (1+\kappa^2)y - 1 + P^2 e^{-2(\phi/\kappa+\xi)} }{y(1-2z)}\,.
\eea
\end{subequations}
As in the purely electric case, the expression for $\dot{\gamma}$ follows from the other equations as long as $\gamma\neq-\kappa$.

The equations depend explicitly on the function $\phi(\xi)$ (instead of just its derivative), which effectively replaces the degree of freedom described by $\omega(\xi)$ in the purely electric case. In spite of this, we can obtain an autonomous system formally identical to the purely electric case~\eqref{eq:AUTO} simply by replacing $Pe^{-(\phi(\xi)/\kappa+\xi)} \to \omega(\xi)$. The analysis of Section~\ref{sec:SphericalCSS} can then be immediately adapted to the case of a purely magnetic Maxwell field.

%%%%%%%%%%%%%%%%%%%%%%%%%%%%%%%%%%

%%%%%%%%%%%%%%%%%%%%%%%%%%%%%%%%%%

\end{document}